\documentclass[prd,aps,nofootinbib]{revtex4-2}

\usepackage{amsmath}
\usepackage{amssymb} 
\usepackage{amsfonts}
\usepackage{booktabs}
\usepackage{hyperref}
\usepackage{slashed} 
\usepackage{graphicx}
\usepackage{diagbox}
\usepackage{float}
\usepackage{color}
\usepackage{makecell}
\usepackage{multirow}
\usepackage{array,tabularx}

\hypersetup{colorlinks=true, linkcolor=blue, citecolor=blue} 

\begin{document}
\title{Strong Decays of the $2^-$ Heavy-Light Mesons in the Relativistic
Bethe-Salpeter Approach}

\author{Jie Yu$^{1,2,3}$,
    Wan-Li Ju$^{1,2,3,4}$,  
    Qiang Li$^5$,
    Tianhong Wang$^6$,
	and Guo-Li Wang$^{1,2,3}$\footnote{Corresponding author}}
\affiliation{$^1$ Department of Physics, Hebei University, Baoding 071002, China\\
	$^2$ Hebei Key Laboratory of High-precision Computation and
	Application of Quantum Field Theory, Baoding 071002, China\\
	$^3$ Hebei Research Center of the Basic Discipline for Computational Physics, Baoding 071002, China\\
    $^4$ Department of Physics, University of Alberta, Edmonton AB T6G 2J1, Canada\\    
    $^5$ School of Physical Science and Technology, Northwestern Polytechnical University, Xi'an 710072, China\\
    $^6$ School of Physics, Harbin Institute of Technology, Harbin 150001, China
    }

\begin{abstract}
We present a relativistic study of the mass spectra, the
$^1D_2$--$^3D_2$ mixing angles, and the OZI-allowed two-body strong
decays of the heavy-light $J^P=2^-$ mesons $D_2$, $D'_2$, $D_{s2}$, $D'_{s2}$, $\bar{B}_2$, $\bar{B}'_2$, $\bar{B}_{s2}$, and $\bar{B}'_{s2}$. The mass spectrum and the
meson wave functions are obtained by solving the full instantaneous
Bethe--Salpeter (Salpeter) equation, in which the wave-function
representations are constructed directly in terms of the $J^P$ quantum
numbers rather than the non-relativistic spectroscopic notation
${}^{2S+1}L_J$. For the decay
amplitudes we use the partial conservation of the axial-vector current
(PCAC) together with the low-energy theorem for the
$H\to H'P$ channels, so that the light pseudoscalar meson enters only
through its decay constant, and the effective chiral Lagrangian for the
$H\to H'V$ channels. The total widths of these mesons are estimated by summing over the partial widths of all OZI-allowed strong decay channels.
\end{abstract}
\maketitle

\section{Introduction}
Over the past two decades, particle physics experiments have achieved remarkable progress, with steadily improving precision and rapidly accumulating event samples, leading to the discovery of many new particles \cite{Chen:2022asf,Brambilla:2019esw}, including a wealth of excited states. These discoveries have sparked tremendous interest in both theoretical and experimental communities. 

Among the newly found excited states are several particles with spin-parity $2^-$. For instance, the charmonium state $\psi_2(3823)$ was first observed by Belle in 2013 \cite{Belle:2013ewt} and later confirmed by BESIII \cite{BESIII:2015iqd} and LHCb \cite{LHCb:2020fvo}; it was identified as the first charmonium $D$-wave meson $\psi_2(1^3D_2)$ with quantum numbers $2^{--}$ \cite{Li:2022qhg}. Similarly, bottomonium $\Upsilon_2(1D)$ was discovered by CLEO in 2004 \cite{CLEO:2004npj} and confirmed by BaBar \cite{BaBar:2010tqb}, also identified as a $D$-wave meson $\Upsilon_2(1^3D_2)$ with $J^{PC} = 2^{--}$.

In the heavy-light meson sector, the BaBar collaboration in 2010 first observed the resonance $D_J(2750)$ \cite{BaBar:2010zpy} in inclusive $e^+e^-$ collisions. Later, in 2013, the LHCb collaboration reported the observation of an unnatural-parity state $D_J(2740)$ \cite{LHCb:2013jjb} in the $D^{*+}\pi^-$ channel from $pp$ collisions, confirming its quantum numbers as $2^-$ and measuring its width to be $73.2\,\text{MeV}$. Subsequently, in 2020, further precise experimental measurements \cite{LHCb:2019juy} determined this resonance and assigned it to the $D_2(2740)$ state, yielding a total decay width of $\Gamma = 102\,\text{MeV}$.
The experimental data on the $D_2(2740)$ have now been updated to \cite{ParticleDataGroup:2024cfk}
\begin{equation}
\begin{aligned}
M &= \left( 2747 \pm 8 \right) \,\text{MeV}, \\
\Gamma &= \left( 89 \pm 19 \right) \,\text{MeV}. \\
\end{aligned}
\end{equation}

However, for other $J^P = 2^-$ heavy-light mesons, including the partner of $D_2(2740)$,  the strange charmed mesons $D_{s2}$ and $D'_{s2}$, the bottom mesons $\bar{B}_{2}$ and $\bar{B'}_{2}$, and the strange bottom mesons $\bar{B}_{s2}$ and $\bar{B'}_{s2}$, remain experimentally unobserved to date. As high-lying excited states predicted by the quark model, these heavy-light mesons are of great importance to meson spectroscopy. Their mass spectrum, production, and decay properties all involve non-perturbative QCD, and will thus serve as crucial tests for existing theoretical frameworks, enabling us to verify the applicability of the quark model.

From the theoretical side, the mass spectrum of heavy-light mesons with $J^P = 2^-$ has been investigated within various frameworks, including relativistic bound state equation \cite{Zeng:1994vj}, relativistic quark model \cite{DiPierro:2001dwf}, QCD-motivated relativistic quark model \cite{Ebert:2009ua}, relativistic GI model \cite{Godfrey:2015dva}, linear potential model \cite{Ni:2021pce}, and non-relativistic screened potential model \cite{M:2026uqc}, etc.  
Theoretical predictions consistently show that even the lowest-lying $2^-$ state lies above the relevant mass threshold, implying that all these states undergo OZI-allowed strong decays. Such OZI-allowed strong decays play a crucial role for a given hadron, since the sum of the partial widths of such decay modes provides an estimate of its total width. Moreover, experimentally, these decay channels serve as key modes for determining the properties of the decaying particle. Therefore, numerous theoretical calculations \cite{Lu:2014zua,Song:2015nia,Godfrey:2015dva,Godfrey:2016nwn,Hao:2022ibj} have been performed on the OZI-allowed strong decays of these heavy-light $2^-$ mesons, producing valuable results.

However, we note that most existing theoretical studies are based on the non-relativistic $^3P_0$ model. While some works have incorporated relativistic corrections, these are often insufficient. We have previously shown that relativistic effects for excited states, particularly highly excited ones, are substantial, and non-relativistic treatments can lead to large errors. Given that $2^-$ mesons belong to highly excited states, precise calculations call for a relativistic framework. Therefore, we employ the instantaneous Bethe-Salpeter equation method to compute the mass spectrum and OZI-allowed strong decays of these $2^-$ heavy-light mesons.

Another crucial observable for $2^-$ mesons is the mixing angle. Within the $S$-$L$ coupling scheme for describing and solving heavy-light mesons, the resulting $^1P_1$ and $^3P_1$ states are not physical eigenstates; rather, the physical states are mixed states of $^1P_1$ and $^3P_1$. In conventional theoretical frameworks, the mixing angle is derived from the spin-orbit and tensor components of the interaction potential. However, our approach neither employs $S$-$L$ coupling nor $j$-$j$ coupling; instead, we construct wave functions directly according to the $J^P$ quantum numbers. Consequently, in Ref.~\cite{Li:2018eqc} we introduced a new method that determines the mixing angle via the wave function normalization condition. The present study continues to adopt this method for extracting the mixing angles. Given that the wave functions are relativistic, the mixing angles obtained thereby are more accurate than those derived from the expansion-based interaction potential method.

The Bethe-Salpeter equation \cite{Salpeter:1951sz}, as a relativistic dynamic equation for a two-body bound state, was first proposed by Bethe and Salpeter in 1951, and later received rigorous proof from Gell-Mann and Low. Due to its nature as an integral equation, it is difficult to solve numerically. Salpeter therefore proposed the instantaneous approximation  \cite{Salpeter:1952ib}, leading to the Salpeter equation, which is suitable for a heavy-light and heavy-heavy meson, greatly reduces the complexity of the calculation and has been widely applied \cite{He:2019ify,Hilger:2014nma,Blank:2011ha}.
Based on the $J^P$ quantum numbers of mesons, we previously constructed relativistic wave-function representations for various mesons, then substituted them into the Salpeter equation for numerical solutions, obtaining eigenvalues and relativistic wave functions for a variety of meson species. These wave functions were further applied to various processes, yielding satisfactory results. At the same time, we also found that relativistic corrections are crucial for many processes. So in this paper, we employ the relativistic meson wave functions obtained by solving the full Salpeter equation to calculate and investigate the properties of $2^-$ mesons.

This paper is organized as follows. Sec.~\ref{sec:decay} describes the method for calculating transition amplitudes and gives the relevant formulas. Sec.~\ref{sec:wave} presents the relativistic wave functions of the mesons involved. Sec.~\ref{sec:results} contains results and discussions, and Sec.~\ref{sec:summary} summarizes the paper.

\section{Formalism of strong decays}\label{sec:decay}
For a heavy-light meson with $J^P = 2^-$, its two-body OZI-allowed strong decays generally lead to a final state consisting of a heavy meson and a light meson. 
Depending on the nature of the final-state light meson, the corresponding decay amplitudes can be classified into two categories: 
one where the light meson is a pseudoscalar meson, and the other where it is a vector meson. 
The former corresponds to decays of the type $H \to H' + P$, while the latter corresponds to $H \to H' + V$. 
We will present the decay amplitude formulas for these two cases separately.
\subsection{$H\to H'+P$}
OZI-allowed strong decay processes generally have branching ratios much larger than those of OZI-suppressed strong decays, as well as electromagnetic and weak decays. 
As they typically constitute the dominant decay channels of a hadron, these processes have attracted considerable attention, and various theoretical approaches have been developed to calculate them. 
One of the most well-known is the $^3P_0$ model \cite{Micu}, in which the transition amplitude is expressed as an overlap integral of the wave functions of the initial and final-state mesons. 
However, we note that the $^3P_0$ model is inherently a non-relativistic framework; consequently, it may induce substantial systematic uncertainties when the final state contains a light meson. 
Therefore, instead of adopting the $^3P_0$ model, we employ the Partial Conservation of the Axial Vector Current (PCAC) relation and utilize low-energy theorem to recast the transition amplitude as an overlap integral of the wave functions of the initial and final heavy mesons \cite{Chang:2005sd}. 
In this scheme, the contribution of the light meson is reduced to its decay constant, thereby allowing us to circumvent the use of a light-meson wave function. This approach was first adopted in Ref.~\cite{Chang:2005sd}, and has since been applied to various OZI-allowed strong decay processes with satisfactory results \cite{Wang:2016enc,Li:2017zng,Liu:2022kvs,Wang:2022tuy}.
In this work, we employ this method to calculate the process $H\to H'+P$, where $P$ is a light pseudoscalar meson.

\begin{figure}[htbp]
    \centering
    \includegraphics[width=0.40\textwidth]{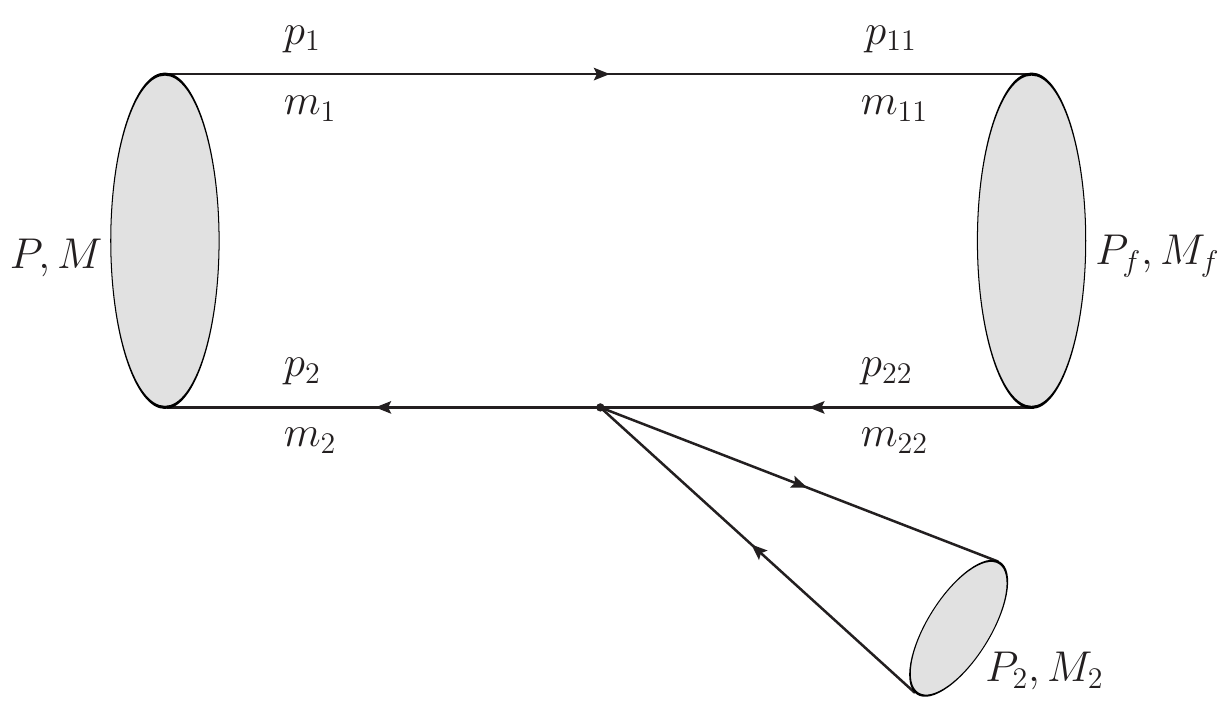}
    \caption{Feynman diagram of the OZI-allowed two-body strong decay}
    \label{fig:feynman}
\end{figure}

Taking $D_2^0 \to D^{*\,+} \pi^-$ as an example, we introduce our calculation method. The corresponding Feynman diagram for this process is given in Fig.~\ref{fig:feynman}. In Fig.~\ref{fig:feynman}, we label the momenta and masses of the mesons, as well as the symbols for the momenta and masses of the internal quarks and antiquarks. In addition, we give the definition of the internal relative momentum. Taking the initial-state meson as an example, we have: $$p_1=\frac{m_1}{m_1+m_2}P+q,\quad
p_2=\frac{m_2}{m_1+m_2}P-q.$$ 
Since we will adopt the instantaneous approximation, we define: $$q_{_{\perp}}= q-q_{_P}\frac{P}{M},~~~~
q_{_P}=\frac{P\cdot q}{M}.$$ 
In the center-of-mass frame of the meson, we have {$P=(M,0)$}, $q_{_{\perp}}=(0,\vec q)$, and $q_{_P}=q_{_0}$.

Applying the PCAC relation and the low-energy theorem, the transition amplitude can be written as \cite{Chang:2005sd}
\begin{equation}\label{amplitude}
\begin{aligned}
\mathcal{M} = \frac{-iP_{2}^\mu}{f_{\pi}}
\langle D^{*+}(P_{f})|\bar{d}\gamma_\mu\gamma_5 u|D_{2}^0(P)\rangle,
\end{aligned}
\end{equation}
where
$f_{\pi}$ is the decay constant of \(\pi\) meson, \(P\), \(P_{f}\), and \(P_2\) are the momenta of  \(D_{2}^{0}\), \(D^{*+}\), and \(\pi\) mesons, respectively. Obviously, apart from its momentum $P_2$, the contribution of the pion meson is manifested in the form of a decay constant, and its decay constant can be fitted experimentally. In this way, we do not need to introduce its wave function as we do for heavy mesons.

It can be seen from Eq. (\ref{amplitude}) that the main calculation now reduces to the computation of the hadronic transition matrix element $\langle D^{*+}(P_{f})|\bar{d}\gamma_\mu\gamma_5 u|D_{2}^{0}(P)\rangle$ between two heavy mesons. Since both are heavy mesons, this hadronic transition matrix element is much easier and more accurate to handle than in the case involving light mesons. Moreover, there are many methods available. We adopt the Mandelstam mechanism \cite{Mandelstam:1955sd}, expressing the hadronic transition matrix element as an overlap integral of the initial and final meson wave functions, namely \cite{Chang:2006tc}:
\begin{equation}
\begin{aligned}
\langle D^{*+}(P_{f})|\bar{d}\gamma_\mu\gamma_5 u|D_{2}^{0}(P)\rangle = \int \frac{d^3\vec{q}}{(2\pi)^3}
\mathrm{Tr}\left[
\bar{\varphi}_{P_{f}}^{++}\left(\vec{q} - \frac{m_{11}}{m_{11} + m_{22}}\vec{P}_f\right)
\frac{\slashed{P}}{M}
\varphi_{P}^{++}(\vec{q}) \gamma_\mu\gamma_5
\right],
\end{aligned}
\label{matrix element}
\end{equation}
where \(\varphi_{P}(\vec{q})\) and \(\varphi_{P_f}(\vec{q_f})\) are the wave functions of initial and final heavy mesons, which are obtained numerically by solving the instantaneous Bethe-Salpeter equation;
\(\vec{q}\) and \(\vec{q_f}\) are the three dimensional relative momentum between quark and antiquark of the initial and final mesons, respectively; As a spectator, $p_1 = p_{11}$, so we have the relation $\vec{q_f}=\vec{q} - \frac{m_{11}}{m_{11} + m_{22}}\vec{P}_f$. In our calculation, since being very small, the contributions of the negative-energy wave functions are neglected, and only the contributions of the positive-energy wave functions \(\varphi_{P}^{++}\) and \(\bar{\varphi}_{P_f}^{++}\) are retained, with, \(\bar{\varphi}_{P_f}^{++}\) = \(\gamma_{0} (\varphi_{P_f}^{++})^+ \gamma_{0}\).

In the case where the final-state light meson is a pseudoscalar, there is a special particle, namely the $\eta$. This is because theoretically what we are familiar with are $\eta_8$ and $\eta_0$, while the physical states are the $\eta$ and $\eta'$ mixed from them:
\begin{equation}
\begin{bmatrix}
\eta \\
\eta'
\end{bmatrix}
=
\begin{bmatrix}
\cos\theta_\eta & \sin\theta_\eta \\
-\sin\theta_\eta & \cos\theta_\eta
\end{bmatrix}
\begin{bmatrix}
\eta_8 \\
\eta_0
\end{bmatrix},
\end{equation}
where the mixing angle $\theta_{\eta} = 19^\circ$ is chosen \cite{Chang:2005sd}.
Therefore, for the strong decay process $D_{s2}^+ \to D_s^{*+} \eta$, the expression for its transition amplitude is \cite{Chang:2005sd}:
\begin{equation}
\begin{aligned}
\mathcal{M} &=i P_{2}^\mu \left[
\frac{-2 M_\eta^2 \cos\theta_\eta}{\sqrt{6} M_{\eta_8}^2 f_{\eta_8}}
+ \frac{M_\eta^2 \sin\theta_\eta}{\sqrt{3} M_{\eta_0}^2 f_{\eta_0}}
\right]\times \langle D_s^{*+}(P_{f})|\bar{s}\gamma_\mu\gamma_5 s|D_{s2}^+(P)\rangle.
\end{aligned}
\end{equation}
The calculation of the hadronic transition matrix element $\langle D_s^{*+}(P_{f})|\bar{s}\gamma_\mu\gamma_5 s|D_{s2}^+(P)\rangle$ is the same as that for the $\langle D^{*+}(P_{f})|\bar{d}\gamma_\mu\gamma_5 u|D_{2}^{0}(P)\rangle$ given above, so it will not be repeated here.

\subsection{$H\to H'+V$}
When the final-state light meson is not a pseudoscalar but a vector meson, such as $\rho$, $\omega$, and $K^*$ etc., PCAC is no longer applicable, and we must adopt other methods to calculate the transition amplitude. We note that in the chiral quark model, the low-energy quark-pseudoscalar meson interaction can be described in the SU(3) flavor basis by the following effective Lagrangian: \cite{Li:1994cy,Li:1997gd,Zhao:2002id,Matsuki:2011xp},
\begin{equation}
\mathcal{L}_{q q P} = \frac{1}{\sqrt{2} f_{\pi}} \, \bar{q}_i \gamma_\mu \gamma_5 q_j \, \partial^\mu \phi_{i j},
\end{equation}
where
\begin{equation}
\phi_{ij} = \sqrt{2}
\begin{bmatrix}
\frac{1}{\sqrt{2}} \pi^0 + \frac{1}{\sqrt{6}} \eta & \pi^+ & K^+ \\
\pi^- & -\frac{1}{\sqrt{2}} \pi^0 + \frac{1}{\sqrt{6}} \eta & K^0 \\
K^- & \bar{K}^0 & -\frac{2}{\sqrt{6}} \eta
\end{bmatrix}
\end{equation}
represents the chiral field of pseudoscalar mesons. 

Starting from this Lagrangian, we obtain a formula for calculating the transition amplitude that is exactly the same as Eq. (\ref{amplitude}). That is to say, the result we obtain by adopting PCAC and the low-energy theorem is the same as that from the low-energy SU(3) theory. Therefore, when the light meson is a vector meson, we can also adopt the effective Lagrangian of the low-energy SU(3) theory to obtain our transition amplitude formula.

In the chiral quark model, the effective Lagrangian for the low-energy quark-vector meson interaction is \cite{Zhao:1998fn,Zhong:2009sk}:
\begin{equation}\label{Vla}
\mathcal{L}_{qqV} = \bar{q}_i \left( a \gamma_\mu + \frac{ib}{2m_j} \sigma_{\mu\nu} P_2^\nu \right) V^\mu_{ij} q_j,
\end{equation}
where $m_j$ is the average of the masses of the two constituent quarks of the light meson, $m_j =  (m_{q_j} + m_{\bar{q}_{i}})/2 $; parameters $a=-3$ and $b=2$ represent the vector and tensor coupling strength, respectively;  $V^{\mu}_{ij}$ is the light vector meson field, 
\begin{equation}
V_{ij} = 
\begin{bmatrix}
\frac{1}{\sqrt{2}} \rho^0 + \frac{1}{\sqrt{2}} \omega & \rho^+ & K^{*+} \\
\rho^- & -\frac{1}{\sqrt{2}} \rho^0 + \frac{1}{\sqrt{2}} \omega & K^{*0} \\
K^{*-} & \bar{K}^{*0} & \phi
\end{bmatrix}.
\end{equation}

Directly adopting the Lagrangian of Eq. (\ref{Vla}), for the process $H\to H'+V$, i.e., when the final-state light hadron is a vector meson $V$, we obtain the formula for calculating the transition amplitude \cite{Tan:2022idi}:
\begin{equation}\label{Vamp}
\mathcal{M} = i \int \frac{d^3 \vec{q}}{(2\pi)^3}
\mathrm{Tr}\left[
\bar{\varphi}_{P_{f}}^{++}(\vec{q} - \frac{m_{11}}{m_{11} + m_{22}}\vec{P}_f)
\frac{\slashed{P}}{M}
\varphi_{P}^{++}(\vec{q})
\left( a\gamma_\mu + \frac{ib}{2m_j} \sigma_{\mu\nu} P_{2}^\nu \right)
\varepsilon_2^\mu
\right],
\end{equation}
consistent with the above, $P$ and $P_f$ are the momenta of the initial and final heavy mesons, 
$P_2$ is the momentum of the final-state light vector meson, and $\varepsilon_2^\mu$ is the polarization vector of the final-state light vector meson. 

\subsection{The Form Factors}
After performing the traces for Eqs.~(\ref{matrix element}) and (\ref{Vamp}) and integrating over the internal relative momentum $\vec{q}$, the transition amplitudes for $H\to H'+P$ and $H\to H'+V$ are expressed in the form of form factors. We still use some typical processes as examples to give their specific expressions, for instance:
\begin{align}
&\mathcal{M}_{D_2 \to D^* \pi}
= \frac{-i}{f_\pi} \left(
\epsilon^{\mu\nu} P_{f\mu} \epsilon_{f\nu} t_1
+ \epsilon^{\mu\nu} P \cdot\epsilon_f P_{f\mu} P_{f\nu} t_2
\right), \label{eq:16} \\[6pt]
&\mathcal{M}_{D_{2} \to D \rho}
= \epsilon^{\mu\nu} P_{f\mu} \epsilon_{2\nu} t_3
+ \epsilon^{\mu\nu} P_{f\mu} P_{f\nu} P_f \cdot \epsilon_2 t_4
+ \epsilon^{\mu\nu} P_{f\mu} P_{f\nu} P \cdot \epsilon_2 t_5, \label{eq:17} \\[6pt]
&\mathcal{M}_{D_2 \to D_2^*(2460) \pi}
= \frac{-i}{f_\pi}\epsilon^{\mu\nu}\epsilon_f^{\alpha\beta}(P_\alpha P_\beta P_{f\mu}P_{f\nu}t_6+P_\beta P_{f\mu}g_{\alpha\nu}t_7+g_{\alpha\mu}g_{\beta\nu}t_8), \label{eq:18} \\[6pt]
&\mathcal{M}_{D_2 \to D_1(2420) \pi}
= \frac{-i}{f_\pi} \epsilon^{\mu\nu} \epsilon_f^\eta \epsilon_{\alpha\beta\nu\eta}
P^{\alpha} P_f^\beta P_{f\mu} t_9, \label{eq:19} \\[6pt]
&\mathcal{M}_{D_2 \to D_0^*(2300) \pi}
= \frac{-i}{f_\pi} \epsilon^{\mu\nu} P_{f\mu} P_{f\nu} t_{10}, \label{eq:20}
\end{align}
where, $\epsilon^{\mu\nu}$ is the polarization tensor of the initial $2^-$ heavy meson; $\epsilon_f^{\mu}$ is the polarization vector of the final heavy $1^-$ or $1^+$ meson; $\epsilon_f^{\alpha\beta}$ is the polarization tensor of the final $2^+$ heavy meson and $\epsilon_2$ is the polarization vector of the final-state light vector meson; $t_i$ $(i=1, 2, \ldots, 10)$ represent the form factors.

\section{Relativistic wave functions}\label{sec:wave}

\subsection{Wave function of $2^-$ meson}
In our calculations, the full Salpeter equation is solved to obtain the wave functions. Despite the use of the instantaneous approximation, meaning that the Salpeter equation and the derived wave functions are not strictly fully relativistic, we retain relativistic information as comprehensively as possible. Specifically, rather than constructing meson wave function representations using the conventional non-relativistic spectroscopic notation ${}^{2S+1}L_J$, we instead employ the $J^P$ quantum numbers. Consequently, the obtained wave functions encapsulate abundant relativistic information. For example, for a $2^-$ meson, under the instantaneous approximation, its general wave function representation is as follows \cite{Wang:2022cxy}:
\begin{equation}
\begin{aligned}\label{wave}
\varphi_{2^{-}}(q_{\perp})
&=\epsilon_{\mu\nu}q_{\perp}^{\mu}q_{\perp}^{\nu}
\left( f_1 + f_2\frac{\slashed{P}}{M} + f_3\frac{\slashed{q}_{\perp}}{M} + f_4\frac{\slashed{P}\slashed{q}_{\perp}}{M^2} \right)\\
&\quad+\frac{i}{M}\epsilon_{\mu\nu\alpha\beta}\gamma^{\mu}P^{\nu}q_{\perp}^{\alpha}\epsilon^{\beta\delta}q_{\perp\delta}
\left( v_1 + v_2\frac{\slashed{P}}{M} + v_3\frac{\slashed{q}_{\perp}}{M} + v_4\frac{\slashed{P}\slashed{q}_{\perp}}{M^2} \right),
\end{aligned}
\end{equation}
where \(\epsilon_{\mu\nu}\) is the polarization tensor of the $2^-$ meson; \(\epsilon_{\mu\nu\alpha\beta}\) is the rank-4 totally antisymmetric tensor, also called the Levi-Civita symbol; $f_i$ and $v_i$ ($i=1,2,3,4$) are the unknown radial wave functions to be solved, and their numerical solutions are obtained by solving the full Salpeter equation. It can be verified that each term in Eq.~(\ref{wave}) carries the quantum numbers $2^-$.
According to the Salpeter equation, the eight radial wave functions are not all independent; only four of them are independent, and they satisfy the following relations \cite{Wang:2022cxy}: 
\begin{equation}
f_3=-f_1Mx_-,\quad f_4=-f_2Mx_+,\quad v_3=v_1Mx_-,\quad v_4=v_2Mx_+,
\end{equation}
where $x_+=(\omega_1+\omega_2)/(m_1\omega_2+m_2\omega_1)$ and $x_-=(\omega_1-\omega_2)/(m_1\omega_2+m_2\omega_1)$; $\omega_i=\sqrt{m^2_i+\vec{q}^2}$ ($i=1,2$) is the quark energy.

We point out that, analogous to the ${}^{1}P_{1}{-}^{3}P_{1}$ mixing of two $1^+$ mesons, there also exist two $2^-$ mesons which are ${}^{1}D_{2}{-}^{3}D_{2}$ mixed states:
\begin{equation}
\begin{aligned}\label{nrmixing}
|D_2\rangle&=|\frac{5}{2}\rangle=\cos\theta |^1D_2\rangle+\sin\theta|^3D_2\rangle,\\
|D_2^{\prime}\rangle&=|\frac{3}{2}\rangle=-\sin\theta |^1D_2\rangle+\cos\theta|^3D_2\rangle.
\end{aligned}
\end{equation}
We note that, in Eq.~(\ref{wave}), the first four terms carry the quantum numbers $2^{-+}$ when $m_1 = m_2$, and they reduce to the ${}^1D_2$ state under non-relativistic conditions (with the $f_3$ and $f_4$ terms neglected). The last four terms yield a $2^{--}$ state for equal-mass constituents and reduce to the ${}^3D_2$ state in the non-relativistic limit (discarding the $v_3$ and $v_4$ terms). It can therefore be seen that the wave functions we provide are all ${}^1D_2-{}^3D_2$ mixed states in the non-relativistic regime.

Unlike conventional approaches, which calculate the mixing angle through the interaction potential, we determine the mixing angle directly from the wave functions.
The normalization condition for the $2^-$ wave function is
\begin{equation}\label{2-nor}
\int \frac{d{\vec q}}{(2\pi)^3}\frac{8
\omega_1\omega_2{\vec q}^4}{15M(m_1\omega_2+m_2\omega_1)}(2f_1f_2-3v_1v_2)=1\equiv \rm{cos}^2\theta+\rm{sin}^2\theta,
\end{equation}
where in the second equation, we have defined the mixing angle $\theta$ between ${}^1 D_2$ and ${}^3D_2$ \cite{Wang:2022cxy}. 
And the numerical value of mixing angle $\theta$ can be calculated from the above definition. 

Based on the Salpeter equation and the $2^-$ wave function Eq. (\ref{wave}), we present the positive-energy wave functions required for calculating the transition matrix element:
\begin{equation}
\begin{aligned}\label{wave++}
\varphi^{++}_{2^{-}}(q_{\perp})
&=\epsilon_{\mu\nu}q_{\perp}^{\mu}q_{\perp}^{\nu}
\left( A_1 + A_2\frac{\slashed{P}}{M} + A_3\frac{\slashed{q}_{\perp}}{M} + A_4\frac{\slashed{P}\slashed{q}_{\perp}}{M^2} \right)\\
&\quad+\frac{i}{M}\epsilon_{\mu\nu\alpha\beta}\gamma^{\mu}P^{\nu}q_{\perp}^{\alpha}\epsilon^{\beta\delta}q_{\perp\delta}
\left( A_5 + A_6\frac{\slashed{P}}{M} + A_7\frac{\slashed{q}_{\perp}}{M} + A_8\frac{\slashed{P}\slashed{q}_{\perp}}{M^2} \right).
\end{aligned}
\end{equation}
The explicit forms of $A_i~(i=1,...,8)$, as functions of the radial wave functions $f_1$, $f_2$, $v_1$, and $v_2$, are provided in Eq.~(\ref{2-}) in Appendix.

As noted above, ${}^1D_2$ and ${}^3D_2$ are non-relativistic representations. Our relativistic wave functions, Eq.~(\ref{wave}), reduce to a mixture of ${}^1D_2$ and ${}^3D_2$ in the non-relativistic limit. The same holds for the positive-energy wave functions in Eq. (\ref{wave++}): terms $A_1$ and $A_2$ correspond to ${}^1D_2$, while terms $A_5$ and $A_6$ correspond to ${}^3D_2$. These four terms are all $D$-waves and give the non-relativistic contributions. In contrast, terms $A_3$, $A_4$, $A_7$, and $A_8$ are relativistic correction terms; they are not $D$-waves but rather $F$-waves. In Ref.~\cite{Wang:2022cxy}, we provided a simple but not rigorous criterion for distinguishing them, namely counting the number of ${q}_{\perp}$ factors. For instance, the terms containing $A_1$, $A_2$, $A_5$, and $A_6$ each contain two ${q}_{\perp}$, corresponding to $D$-waves; whereas the remaining terms each contain three ${q}_{\perp}$, corresponding to $F$-waves.

\subsection{Wave functions of $0^-$ and $1^-$ mesons}
Since it is not the wave functions themselves but their corresponding positive-energy wave functions that appear in Eq.~(\ref{Vamp}) for calculating the transition matrix elements, for other types of particles involved in this work we do not present their wave functions; instead, we directly  list  their positive-energy wave functions. Readers interested in the wave-function representations may find them in the references provided below. For pseudoscalar ($0^-$) and vector ($1^-$) mesons, such as $D$, $\bar{B}$, $D^*$, and $B^*$, their positive-energy wave functions are expressed as
 \cite{Ke:2023mrt,Liu:2022kvs},
\begin{equation}
\begin{aligned}\label{positive energy wave 0-}
\varphi_{0^{-}}^{++}(q_{\perp})=
\left( B_1 + B_2\frac{\slashed{P}}{M} + B_3\frac{\slashed{q}_{\perp}}{M} + B_4\frac{\slashed{P}\slashed{q}_{\perp}}{M^2} \right)\gamma^{5},
\end{aligned}
\end{equation}
and
\begin{equation}
\begin{aligned}\label{positive energy wave 1-}
\varphi_{1^{-}}^{++}(q_{\perp})
&=q_{\perp}\cdot\epsilon\left(C_1+ C_2\frac{\slashed{P}}{M}+C_3\frac{\slashed{q}_{\perp}}{M}+C_4\frac{\slashed{P}\slashed{q}_{\perp}}{M^2} \right)\\
&\quad+M\slashed{\epsilon}\left(C_5+ C_6\frac{\slashed{P}}{M}+C_7\frac{\slashed{q}_{\perp}}{M}+C_8\frac{\slashed{P}\slashed{q}_{\perp}}{M^2} \right),
\end{aligned}
\end{equation}
where the coefficients $B_i~(i=1,2,...,4)$ and $C_i~(i=1,2,...,8)$ are functions of the radial wave functions, whose explicit expressions are given in Eqs.~(\ref{0-}) and (\ref{1-})  in the Appendix.

In Eq.~(\ref{positive energy wave 0-}), the terms containing $B_1$ and $B_2$ do not involve $q_\perp$, that is, they contain zero $q_\perp$ factors. According to Ref.~\cite{Wang:2022cxy}, these correspond to ${}^1S_0$ and are $S$-waves, providing the non-relativistic contributions. By contrast, the terms containing $B_3$ and $B_4$ each contain one $q_\perp$ factor; they are not $S$-waves but rather $P$-waves, providing relativistic correction contributions. 

The relativistic wave function of the $1^-$ vector meson is relatively complex, containing eight terms, as detailed in Eq.~(\ref{positive energy wave 1-}). The $C_5$ and $C_6$ terms are free of $q_\perp$ and hence represent $S$-waves, which yield the non-relativistic contributions and correspond to ${}^3S_1$. All other terms are relativistic correction. Specifically, terms $C_1$, $C_2$, $C_7$, and $C_8$ each contain a single $q_\perp$ factor and are therefore $P$-waves. Terms $C_3$ and $C_4$ contain two $q_\perp$ factors, with the explicit structure $q_\perp\!\cdot\!\epsilon{\slashed{q}_\perp}$; nevertheless, these cannot be straightforwardly assigned to $D$-waves, since they also supply an $S$-wave portion. The $D$-wave part is given by $q_\perp\!\cdot\!\epsilon{\slashed{q}_\perp} - \frac{1}{3}q_\perp^2{\slashed{\epsilon}}$, whereas the $S$-wave part reads $\frac{1}{3}q_\perp^2{\slashed{\epsilon}}$.

\subsection{Wave functions of $0^+$, $1^+$ and $2^+$ mesons}
For the $0^+$, $1^+$ and $2^+$ mesons, their positive energy wave functions can be written as \cite{Tan:2019hbw,Li:2017zng,Tan:2022idi},
\begin{equation}
\begin{aligned}\label{0+w}
\varphi_{0^{+}}^{++}(q_\perp)
=G_1+ G_2\frac{\slashed{P}}{M}+G_3\frac{\slashed{q}_{\perp}}{M}+G_4\frac{\slashed{P}\slashed{q}_{\perp}}{M^2},
\end{aligned}
\end{equation}
\begin{equation}
\begin{aligned}\label{1+w}
\varphi_{1^{+}}^{++}(q_\perp)
&=q_{\perp}\cdot\epsilon\left(E_1+ E_2\frac{\slashed{P}}{M}+E_3\frac{\slashed{q}_{\perp}}{M}+E_4\frac{\slashed{P}\slashed{q}_{\perp}}{M^2} \right)\gamma^5\\
& + i \varepsilon_{\mu \nu \alpha \beta} 
\frac{P^\nu}{M} q_\perp^\alpha \epsilon^\beta \gamma^\mu
\left(
E_5+ \frac{\slashed{P}}{M} E_6 + \frac{\slashed{q}_\perp}{M} E_7
+ \frac{\slashed{P} \slashed{q}_\perp}{M^2} E_8
\right),
\end{aligned}
\end{equation}
and
\begin{equation}
\begin{aligned}\label{2+w}
\varphi_{2^{+}}^{++}(q_\perp)
&=\epsilon_{\mu\nu}q_\perp^{\mu}q_\perp^{\nu}\left(F_1+ F_2\frac{\slashed{P}}{M}+F_3\frac{\slashed{q}_{\perp}}{M}+F_4\frac{\slashed{P}\slashed{q}_{\perp}}{M^2} \right)\\
&+M\epsilon_{\mu\nu}\gamma^{\mu}q_\perp^{\nu}\left(F_5+ F_6\frac{\slashed{P}}{M}+F_7\frac{\slashed{q}_{\perp}}{M}+F_8\frac{\slashed{P}\slashed{q}_{\perp}}{M^2} \right),
\end{aligned}
\end{equation}
where the expressions of $G_i~(i=1,2,...,4)$, $E_i~(i=1,2,...,8)$ and $F_i~(i=1,2,...,8)$ are shown Eqs.~(\ref{0+}), (\ref{1+}) and (\ref{2+})  in the Appendix.

For the $0^+$ state wave function given in Eq.~(\ref{0+w}), the non-relativistic contributions originate from terms $G_3$ and $G_4$, which contain a single $q_\perp$ factor and represent $P$-waves corresponding to ${}^3P_0$. In contrast, terms $G_1$ and $G_2$ are free of $q_\perp$ and thus constitute $S$-waves that yield relativistic corrections. 

In the $1^+$ state wave function, Eq.~(\ref{1+w}), terms $E_1$ and $E_2$ each involve one $q_\perp$ factor; they are $P$-waves corresponding to ${}^1P_1$. Terms $E_5$ and $E_6$ are also $P$-waves, corresponding to ${}^3P_1$. These $P$-wave terms furnish the non-relativistic contributions, whereas the remaining terms, each containing two $q_\perp$ factors, are $D$-waves supplying relativistic corrections. 

A parallel can be drawn between the $2^+$ state wave function in Eq.~(\ref{2+w}) and that of the $1^-$ state in Eq. (\ref{positive energy wave 1-}). Specifically, terms $F_5$ and $F_6$, containing one $q_\perp$ factor, are $P$-waves that provide non-relativistic contributions and correspond to ${}^3P_2$. The rest are relativistic corrections: terms $F_1$, $F_2$, $F_7$, and $F_8$ contain two $q_\perp$ factors and are $D$-waves; terms $F_3$ and $F_4$ contain three $q_\perp$ factors and cannot be simply treated as $F$-waves, since they also deliver a partial $P$-wave contribution. The $F$-wave part is $(\epsilon_{\mu\nu}q_\perp^{\mu}q_\perp^{\nu} {\slashed{q}_\perp} - \frac{2}{5}q_\perp^2 \epsilon_{\mu\nu}{\gamma}^{\mu}q_\perp^{\nu})$, and the $P$-wave part is $\frac{2}{5}q_\perp^2 \epsilon_{\mu\nu}{\gamma}^{\mu}q_\perp^{\nu}$.

\section{Results and discussions}\label{sec:results}
In this section, we give our results and compare ours with those from other models. In our method, the following parameters are used: $m_c = 1.62\,\text{GeV}$, 
$m_b = 4.96\,\text{GeV}$, 
$m_u =m_d=m_q= 0.31\,\text{GeV}$, 
$m_s = 0.50\,\text{GeV}$, $m_{\eta_8} = 0.5643\,\text{GeV}$, and $m_{\eta_0} = 0.9481\,\text{GeV}$. 
The decay constants used in this work are: $f_{\pi} = 130.2\,\text{MeV}$, $f_{K} = 156.2\,\text{MeV}$, $f_{\eta_8}=1.26f_{\pi}$, and $f_{\eta_0}=1.07f_{\pi}$. 

\subsection{Mass spectra and mixing angles of $2^-$ mesons}
We solve the full Salpeter equation fulfilled by the $2^-$ meson wave functions to obtain numerical solutions of the radial wave functions and the resulting mass spectrum, which is displayed in Table~\ref{mixing}. In our framework, because wave function representations for distinct $J^P$ quantum numbers are given independently, the mass $M_l$ of the lighter $2^-$ state serves as an input parameter, whereas the mass $M_h$ of the heavier state is predicted. The mass difference between them more sensitively reveals differences among various model approaches; hence $\Delta M=M_{h}-M_l$ is also tabulated in Table~\ref{mixing}. With the wave functions at hand and employing Eq.~(\ref{2-nor}), we compute the ${}^{1}D_{2}{-}^{3}D_{2}$ mixing angle $\theta$, with the results likewise listed in Table~\ref{mixing}. For completeness, selected results from other theoretical works are also quoted in Table~\ref{mixing}. We note that Refs.~\cite{Zeng:1994vj,DiPierro:2001dwf,Liu:2016efm} employ heavy quark effective theory, or equivalently the $j$-$j$ coupling scheme, where the concept of a mixing angle does not apply.

\begin{table}[h]
\centering
\caption{Mixing angles, masses (MeV), and mass differences of the ${}^{1}D_{2}{-}^{3}D_{2}$ mixed $2^-$ meson states.}
\vspace{3mm}
\setlength{\tabcolsep}{12pt}
\begin{tabular}{ccccccccc}
\hline
 & &ours&\cite{Zeng:1994vj}&\cite{DiPierro:2001dwf}&\cite{Ebert:2009ua}&\cite{Godfrey:2015dva,Godfrey:2016nwn} &\cite{Liu:2016efm}&\cite{li:2021hss,Ni:2021pce} \\
\hline
$c\bar{q}$ & $D_2$      &      $2747$     &2740 &2775 &2806          &2816           &2737 &2755 \\
           & ${D'_2}$ &      $2803$     &2760 &2833 &2850          &2845           &2881 &2827 \\
           &$\Delta M$    &      $56$       &$20$ &$58$ &$44$          &$29$           &144  &72   \\
           &$\theta$      &$59.2^{\circ}$  &--   &--   &$40.7^{\circ}$&$-38.2^{\circ}$&--   &$-40.2^{\circ}$ \\
           \hline
$c\bar{s}$ & $D^+_{s2}$   &      $2900$     &2860 &2900 &2931          &2900           &2857 &2857 \\
           &${D'^{+}_{s2}}$&     $2936$     &2880 &2953 &2961          &2926           &2923 &2911 \\
           &$\Delta M$    &      $36$       &$20$ &$53$ &$30$          &$26$           &66   &54   \\         
           &$\theta$      &$60.5^{\circ}$  &--   &--   &$39.2^{\circ}$&$-38.5^{\circ}$&--   &$-40.7^{\circ}$ \\
           \hline
$b\bar{q}$ & $\bar{B}_2$      &      $6095$     &5960 &5985 &6103          &6095           &5993 &5973 \\ 
           &${\bar{B}'_2}$  &      $6146$     &5980 &6037 &6121          &6124           &6165 &6067 \\ 
           &$\Delta M$    &      $51$       &$20$ &$52$ &$18$          &$29$           &172  &94   \\          
           &$\theta$      &$56.1^{\circ}$  &--   &--   &$38.0^{\circ}$&$39.7^{\circ}$ &--   &$-39.5^{\circ}$ \\
           \hline
$b\bar{s}$&$\bar{B}_{s2}$ &      $6169$     &6070 &6095 &6189          &6169           &6098 &6061 \\
          &${\bar{B'}_{s2}}$&    $6203$     &6080 &6140 &6218          &6196           &6186 &6113 \\
           &$\Delta M$    &      $34$       &$10$ &$45$ &$29$          &$27$           &88   &52   \\         
           &$\theta$      &$56.3^{\circ}$  &--   &--   &$38.1^{\circ}$&$40.0^{\circ}$ &--   &$-39.9^{\circ}$ \\
\hline
\end{tabular}
\label{mixing}
\end{table}

For the $2^-$ charm meson $D_2$, we use the experimental mass $M_{D_{2}} = 2747\,\text{MeV}$ of the $D_2(2740)$ as input, and take the predicted mass from our method, $M_{D'_{2}} = 2803\,\text{MeV}$, for the undiscovered ${D'_2}$ state. The mass splitting between $D_2$ and $D'_2$ is estimated to be $\Delta M = 56\,\text{MeV}$, which is in good agreement with prediction from Ref.~\cite{DiPierro:2001dwf} in Table~\ref{mixing} and is close to the value of Ref.~\cite{Ebert:2009ua}.

For the charm-strange meson $D_{s2}$, we adopt the input mass $M_{D_{s2}} = 2900\,\text{MeV}$ taken from the GI model prediction in Ref.~\cite{Godfrey:2015dva}, and the mass of the high-mass mixed state is obtained as $M(D'_{s2}) = 2936\,\text{MeV}$. The mass splitting $\Delta M = 36\,\text{MeV}$ between the two states is close to the prediction in Ref.~\cite{Ebert:2009ua}.

For $\bar{B}_2$ and $\bar{B}_{s2}$, the mass values are likewise taken from the GI model \cite{Godfrey:2016nwn}, with inputs $M_{\bar{B}_2} = 6095$ MeV and $M_{\bar{B}_{s2}} = 6169$  MeV. 
Our theoretical predictions for the heavier states are $M_{\bar{B}'_2} = 6146$  MeV and $M_{\bar{B}'_{s2}} = 6203$ MeV. 
The mass splitting between $\bar{B'}_2$ and $\bar{B}_2$ is $\Delta M = 51$  MeV, which is nearly twice the GI model value $29$ MeV~\cite{Godfrey:2016nwn}, but close to $52$ MeV in Ref.~\cite{DiPierro:2001dwf}. 
Similarly, the mass splitting between $\bar{B}'_{s2}$ and $\bar{B}_{s2}$ is $\Delta M = 34$  MeV, showing good consistency with $29$ MeV in Ref.~\cite{Ebert:2009ua}.

Table~\ref{mixing} lists our predictions for the mixing angles together with those from other theoretical approaches. It can be seen that the mixing angles predicted by different models differ significantly, with even the sign of the angle varying among them. For our method, it is obvious that Eq. (\ref{2-nor}) cannot give a unique angle with definite value and sign since it is an equation with multiple solutions. However, since we have given the wave functions of $^1D_2$ and $^3D_2$ states separately in Ref. \cite{Li:2016efw}, comparing the numerical values of the full wave function in Eq. (\ref{wave}) for the $2^-$ state with those of the $^1D_2$ wave function in Eq. (C.1) and $^3D_2$ wave function in Eq. (C.3) in Ref. \cite{Li:2016efw}, we can uniquely determine the value and the sign of the mixing angle, and conclude that the definition of the mixing in Eq. (\ref{2-nor}) is equivalent to the commonly used formula in Eq. (\ref{nrmixing}). 

Concerning the issue of differing mixing angle signs, we pointed out in Ref.~\cite{Wang:2022tuy} that this implies that the masses of the two mixed states, namely the lighter and heavier states, are flipped, or that the quark composition changes from $Q\bar{q}$ to $\bar{Q}q$.

\subsection{Strong decays of $D_2$ and $D'_2$}

Tables~\ref{D2} and \ref{D2'} list, respectively, the strong decay results for $D_2$ and $D'_2$ obtained by our group and by other theoretical groups. Our analysis indicates that the $D^*\pi$, $D\rho$, and $D\omega$ channels dominate the decays of $D_2$, contributing $36\%$, $27\%$, and $18\%$, respectively. Although the branching fractions for individual channels vary across different models, our predicted total width of $99.2~\text{MeV}$ agrees well with the theoretical results of $113.34$ and $109.43$ MeV from Refs.~ \cite{Yu:2014dda,Lu:2014zua}, and is also consistent with the experimental measurement of $89 \pm 19~\text{MeV}$.   

It can be seen from Table~\ref{D2'} that the dominant decay channels of the higher-mass state $D'_2$ are distinct from those of the lower-mass state $D_2$. The dominant decay modes are $D'_2 \to D_2^*(2460)\pi$ and $D'_2 \to D^*\pi$, with partial widths of $163$ MeV and $91~\text{MeV}$, and corresponding branching fractions of $56\%$ and $31\%$, respectively. Together, these two channels constitute $87\%$ of the total width of $D'_2$. Our calculated partial width for $D'_2 \to D_2^*(2460)\pi$ agrees very well with the values of $145$ MeV and $148~\text{MeV}$ reported in Refs.~\cite{Lu:2014zua,Hao:2024ptu}. Furthermore, our predicted total width of $291~\text{MeV}$ for $D'_2$ shows good consistency with the results of $244$ MeV, $258.4$ MeV, and $279.9~\text{MeV}$ from Refs.~\cite{Godfrey:2015dva,Ni:2021pce,Lu:2014zua}.

\begin{table}
\caption{Two-body strong decay widths (MeV) of the $D_2(2740)$ meson.}
\vspace{0.2cm}
\setlength{\tabcolsep}{0.01cm}
\centering
\begin{tabular*}{\textwidth}{@{}@{\extracolsep{\fill}}ccccccc}
\hline\hline
Mode & Ours & Ref. \cite{Yu:2014dda} & Ref. \cite{Lu:2014zua} & Ref. \cite{Hao:2024ptu} & Ref. \cite{Ni:2021pce} & Ref. \cite{Godfrey:2015dva} \\
\hline
{\phantom{\Large 1}}\raisebox{+.2cm}{\phantom{\Large j}}
$D^{*+}\pi^-$ & 23.8 & 17.04 & 13.5 & \multirow{2}{*}{58} & \multirow{2}{*}{30.4} & \multirow{2}{*}{21.2} \\
{\phantom{\Large 1}}\raisebox{+.2cm}{\phantom{\Large j}}
$D^{*0}\pi^0$ & 12.3 & 8.74 & 7 & &  &  \\
{\phantom{\Large 1}}\raisebox{+.2cm}{\phantom{\Large j}}
$D^+\rho^-$ & 17.4 & 36.02 & 39.39 & \multirow{2}{*}{118} & \multirow{2}{*}{5.8} & \multirow{2}{*}{61.2} \\
{\phantom{\Large 1}}\raisebox{+.2cm}{\phantom{\Large j}}
$D^0\rho^0$ & 9.23 & 18.90 & 20.78 & &  &  \\
{\phantom{\Large 1}}\raisebox{+.2cm}{\phantom{\Large j}}
$D\omega$ & 17.6 & 16.90 & 19.34 & 38 & 1.7 & 19.6 \\
{\phantom{\Large 1}}\raisebox{+.2cm}{\phantom{\Large j}}
$D_s^{*+}K^-$ & 1.56 & 0.38 & 0.35 & 1 & 0.9 & -- \\
{\phantom{\Large 1}}\raisebox{+.2cm}{\phantom{\Large j}}
$D^{*0}\eta$ & 1.62 & 0.59 & 0.93 & -- & 0.8 & -- \\
{\phantom{\Large 1}}\raisebox{+.2cm}{\phantom{\Large j}}
$D^{*+}_0(2300)\pi^-$ & 4.61 & 0.06 & 0.11 & \multirow{2}{*}{2} & \multirow{2}{*}{8.1} & -- \\
{\phantom{\Large 1}}\raisebox{+.2cm}{\phantom{\Large j}}
$D_0^{*0}(2300)\pi^0$ & 2.31 & -- & 0.07 & &  & -- \\
{\phantom{\Large 1}}\raisebox{+.2cm}{\phantom{\Large j}}
$D_2^{*+}(2460)\pi^-$ & 4.82 & -- & 4.66 & \multirow{2}{*}{1} & \multirow{2}{*}{4.4} & -- \\
{\phantom{\Large 1}}\raisebox{+.2cm}{\phantom{\Large j}}
$D_2^{*0}(2460)\pi^0$ & 2.42 & 0.97 & 2.36 & &  & -- \\
{\phantom{\Large 1}}\raisebox{+.2cm}{\phantom{\Large j}}
$D_1^{+}(2420)\pi^-$ & 0.25 & -- & 0.57 & \multirow{2}{*}{2} & \multirow{2}{*}{5.2} & -- \\
{\phantom{\Large 1}}\raisebox{+.2cm}{\phantom{\Large j}}
$D_1^{0}(2420)\pi^0$ & 0.13 & 0.26 & 0.31 & &  & -- \\
{\phantom{\Large 1}}\raisebox{+.2cm}{\phantom{\Large j}}
$D_1^{+}(2430)\pi^-$ & 0.76 & -- & 0.04 & \multirow{2}{*}{1} & \multirow{2}{*}{0.1} & -- \\
{\phantom{\Large 1}}\raisebox{+.2cm}{\phantom{\Large j}}
$D_1^{0}(2430)\pi^0$ & 0.39 & -- & 0.02 & &  & -- \\
\hline
{\phantom{\Large 1}}\raisebox{+.2cm}{\phantom{\Large j}}
$\Gamma_{total}$ & 99.2 & 113.34 & 109.43 & 221 & 57.4 & 105 \\
{\phantom{\Large 1}}\raisebox{+.2cm}{\phantom{\Large j}}
EXP & \multicolumn{6}{c}{$89\pm19$} \\
\hline\hline
\end{tabular*}
\label{D2}
\end{table}

Finally, a comparison of the total widths of $D_2$ and $D'_2$ predicted by all theoretical models compiled in Tables~\ref{D2} and \ref{D2'} reveals a consistent pattern: the lower-mass state $D_2$ is narrow, whereas the corresponding higher-mass state $D'_2$ is broad. Experimentally, detecting a broad state such as $D'_2$ is challenging, while identifying a narrow state is comparatively easier, a situation that aligns well with the present experimental status.

\begin{table}
\caption{Two-body strong decay widths (MeV) of the $D'_2$ state.}
\vspace{0.2cm}
\setlength{\tabcolsep}{0.01cm}
\centering
\begin{tabular*}{\textwidth}{@{}@{\extracolsep{\fill}}ccccccc}
\hline\hline
Mode & Ours & Ref. \cite{Yu:2014dda} & Ref. \cite{Lu:2014zua} & Ref. \cite{Hao:2024ptu} & Ref. \cite{Ni:2021pce} & Ref. \cite{Godfrey:2015dva} \\
\hline
{\phantom{\Large 1}}\raisebox{+.2cm}{\phantom{\Large j}}
$D^{*+}\pi^-$ & 60.2 & 48.63 & 75.98 & \multirow{2}{*}{46} & \multirow{2}{*}{91.6} & \multirow{2}{*}{93.3} \\
{\phantom{\Large 1}}\raisebox{+.2cm}{\phantom{\Large j}}
$D^{*0}\pi^0$ & 30.6 & 24.24 & 38.21 & & & \\
{\phantom{\Large 1}}\raisebox{+.2cm}{\phantom{\Large j}}
$D^+\rho^-$ & 1.82 & 0.5 & 0.39 & \multirow{2}{*}{30} & \multirow{2}{*}{15.2} & \multirow{2}{*}{5.85} \\
{\phantom{\Large 1}}\raisebox{+.2cm}{\phantom{\Large j}}
$D^0\rho^0$ & 0.98 & 0.29 & 0.23 & & & \\
{\phantom{\Large 1}}\raisebox{+.2cm}{\phantom{\Large j}}
$D\omega$ & 1.79 & 0.2 & 0.18 & 10 & 4.3 & -- \\
{\phantom{\Large 1}}\raisebox{+.2cm}{\phantom{\Large j}}
$D_s^{*+}K^-$ & 16.5 & 6.95 & 6.40 & 42 & 17.5 & 29.5 \\
{\phantom{\Large 1}}\raisebox{+.2cm}{\phantom{\Large j}}
$D^{*0}\eta$ & 12.6 & 7.41 & 13.34 & 0 & 12 & 14.7 \\
{\phantom{\Large 1}}\raisebox{+.2cm}{\phantom{\Large j}}
$D^{*+}_0(2300)\pi^-$ & 1.32 & -- & 0.02 & \multirow{2}{*}{4} & \multirow{2}{*}{0.02} & -- \\
{\phantom{\Large 1}}\raisebox{+.2cm}{\phantom{\Large j}}
$D^{*0}_0(2300)\pi^0$ & 0.71 & 0.00028 & 0.02 & & & -- \\
{\phantom{\Large 1}}\raisebox{+.2cm}{\phantom{\Large j}}
$D_2^{*+}(2460)\pi^-$ & 109 & -- & 96.17 & \multirow{2}{*}{148} & \multirow{2}{*}{117.1} & \multirow{2}{*}{89.9} \\
{\phantom{\Large 1}}\raisebox{+.2cm}{\phantom{\Large j}}
$D_2^{*0}(2460)\pi^0$ & 54.2 & 23.26 & 48.57 & & & \\
{\phantom{\Large 1}}\raisebox{+.2cm}{\phantom{\Large j}}
$D_1^{+}(2420)\pi^-$ & 0.39 & -- & 0.19 & \multirow{2}{*}{2} & \multirow{2}{*}{0.2} & -- \\
{\phantom{\Large 1}}\raisebox{+.2cm}{\phantom{\Large j}}
$D_1^{0}(2420)\pi^0$ & 0.19 & 0.089 & 0.10 & & & -- \\
{\phantom{\Large 1}}\raisebox{+.2cm}{\phantom{\Large j}}
$D_1^{+}(2430)\pi^-$ & 0.88 & -- & 0.05 & \multirow{2}{*}{2} & \multirow{2}{*}{0.5} & -- \\
{\phantom{\Large 1}}\raisebox{+.2cm}{\phantom{\Large j}}
$D_1^{0}(2430)\pi^0$ & 0.44 & -- & 0.03 & & & -- \\
\hline
{\phantom{\Large 1}}\raisebox{+.2cm}{\phantom{\Large j}}
$\Gamma_{total}$ & 291 & 111.56 & 279.90 & 304 & 258.4 & 244 \\
\hline\hline
\end{tabular*}
\label{D2'}
\end{table}

\subsection{Strong decays of $D^+_{s2}$ and $D'^+_{s2}$}

In Table~\ref{Ds2}, we present the decay widths of $D_{s2}$ and $D_{s2}'$ obtained in this work and compare them with results from different models. Refs.~\cite{Godfrey:2015dva,Yang:2023tvc,Hao:2022vwt} all employ the $^3P_0$ model but use different parameter sets, leading to some dispersion in their predictions. 
Because the light quark in $D_{s2}$ and $D'_{s2}$ is a strange quark, the OZI-allowed strong decay final states involve light mesons of the $K$ or $\eta$ type. As a result, phase space suppression restricts the OZI-allowed strong decay modes of $D_{s2}$ and $D'_{s2}$ to far fewer channels compared to $D_2$ and $D'_2$. As evident from Table~\ref{Ds2}, only three decay channels exist.

For the lower-mass $D_{s2}$ state, our calculated total decay width $\Gamma \simeq 98.0\,\text{MeV}$ is close to the result of 115.1 MeV in Ref.~\cite{Godfrey:2015dva} and 104 MeV in Ref.~\cite{Yang:2023tvc}. The dominant decay channels in our method are $DK^*$ and $D^*K$, with partial widths of $45.8\,\text{MeV}$ and $50.2\,\text{MeV}$, respectively.  

For the higher-mass state $D'_{s2}$, our prediction indicates that the dominant decay mode is $D^*K$, with a partial width of $131~\text{MeV}$, which accounts for about $88\%$ of the total width. The total width of $149~\text{MeV}$ we predicted  is relatively close to the value of $134~\text{MeV}$ given in Ref.~\cite{Yang:2023tvc}.

\begin{table}[htbp]
\centering
\caption{Two-body strong decay widths (MeV) of the $D_{s2}$ and $D'_{s2}$ states.}
\vspace{0.2cm}
\setlength{\tabcolsep}{0.2cm}
\begin{tabular*}{\textwidth}{@{}@{\extracolsep{\fill}}ccccccc@{}}
\hline\hline
Mode & ours & Ref. \cite{Godfrey:2015dva} & Ref. \cite{Song:2015nia} & Ref. \cite{Ni:2021pce} & Ref. \cite{Yang:2023tvc} & Ref. \cite{Hao:2022vwt} \\
\hline
\multicolumn{7}{l}{$D_{s2}$} \\
\phantom{\Large{1}}\raisebox{+.2cm}{\phantom{\Large{j}}}
$DK^*$ & $45.8$ & $93.2$ & $\approx 10$ & $13.8$ & $38$ & $99$ \\
\phantom{\Large{1}}\raisebox{+.2cm}{\phantom{\Large{j}}}
$D^*K$ & $50.2$ & $21.1$ & $\approx 210$ & $107.1$ & $62$ & $45$ \\
\phantom{\Large{1}}\raisebox{+.2cm}{\phantom{\Large{j}}}
$D_s^*\eta$ & $2.01$ & $0.7$ & $\approx 20$ & $10.5$ & $4$ & $1$ \\
\phantom{\Large{1}}\raisebox{+.2cm}{\phantom{\Large{j}}}
$\Gamma_{total}$ & $98.0$ & $115.1$ & $240$ & $131.4$ & $104$ & $145$ \\
\hline
\multicolumn{7}{l}{$D'_{s2}$} \\
\phantom{\Large{1}}\raisebox{+.2cm}{\phantom{\Large{j}}}
$DK^*$ & $7.54$ & $7.92$ & $\approx 140$ & $10.6$ & $53$ & $11$ \\
\phantom{\Large{1}}\raisebox{+.2cm}{\phantom{\Large{j}}}
$D^*K$ & $131$ & $163$ & $\approx 7$ & $28.4$ & $75$ & $78$ \\
\phantom{\Large{1}}\raisebox{+.2cm}{\phantom{\Large{j}}}
$D_s^*\eta$ & $9.84$ & $24.6$ & $\approx 0$ & $1.8$ & $6$ & $17$ \\
\phantom{\Large{1}}\raisebox{+.2cm}{\phantom{\Large{j}}}
$\Gamma_{total}$ & $149$ & $195$ & $147$ & $44$ & $134$ & $106$ \\
\hline\hline
\end{tabular*}
\label{Ds2}
\end{table}

A comparison of the decays of the lower-mass $D_{s2}$ and higher-mass $D'_{s2}$ reveals a discrepancy among theoretical calculations regarding which state is broad and which is narrow. Our results, in conjunction with Refs.~\cite{Godfrey:2015dva,Yang:2023tvc}, favor the assignment of the lower-mass $D_{s2}$ as a narrow state and the higher-mass $D'_{s2}$ as a broad state. Conversely, other theoretical studies~\cite{Ni:2021pce,Hao:2022vwt,Song:2015nia} advocate the opposite assignment. As noted above, this situation exemplifies a mass interchange between certain broad and narrow states. In Table~\ref{Ds2}, we designate the lower-mass state as $D_{s2}$ and the higher-mass state as $D'_{s2}$. 
If we categorize the states not according to their masses but rather as narrow or broad, our obtained total width of $98.0~\text{MeV}$ for the narrow state is consistent with the $106~\text{MeV}$ reported in Ref.~\cite{Hao:2022vwt}. Furthermore, our total width of $149~\text{MeV}$ for the broad state is in good agreement with the $145~\text{MeV}$ from Ref.~\cite{Hao:2022vwt}, and is also comparable to the $131.4~\text{MeV}$ result from Ref.~\cite{Ni:2021pce}.

\subsection{Strong decays of $\bar{B}_{2}$ and $\bar{B'}_{2}$}
Tables~\ref{B2} and \ref{B2'} compile the predictions for strong decays of the $2^-$ bottom mesons $\bar{B}_{2}$ and $\bar{B'}_{2}$ obtained by various theoretical models. Based on the total widths presented in both tables, our results and those of Refs.~\cite{Godfrey:2016nwn,Chen:2022fye,Asghar:2018tha} unanimously show that the lower-mass state $\bar{B}_{2}$ is narrow, whereas the higher-mass state $\bar{B'}_{2}$ is broad. Conversely, Refs.~\cite{Lu:2016bbk,li:2021hss} lead to the opposite assignment, namely that the lower-mass $\bar{B}_{2}$ is broad and the higher-mass $\bar{B'}_{2}$ is narrow.

For the narrow state, our analysis indicates that the dominant decay modes are $B^*\pi$ and $B\rho$, with partial widths of $48.5~\text{MeV}$ and $13.6~\text{MeV}$, and corresponding branching fractions of $58\%$ and $16\%$, respectively. While the individual partial widths vary, Refs.~\cite{Godfrey:2016nwn,Lu:2016bbk} independently arrive at the same conclusion: the primary decay channels of the narrow state are $B^*\pi$ and $B\rho$. Our obtained total width for the narrow state is approximately $83.4~\text{MeV}$, which agrees well with the $89.2~\text{MeV}$ reported in Ref.~\cite{li:2021hss}.

For the broad state in the $2^-$ bottom meson sector, our results indicate that the dominant decay modes are $B^*\pi$ and $B_2^*(5747)\pi$, with partial widths of $119~\text{MeV}$ and $249~\text{MeV}$, and corresponding branching fractions of $27\%$ and $57\%$, respectively. While Refs.~\cite{Godfrey:2016nwn,Chen:2022fye,li:2021hss} similarly identify these two channels as the primary ones, there are minor discrepancies in their relative importance. Ref.~\cite{Chen:2022fye} is consistent with our finding that $B_2^*(5747)\pi$ contributes the most, while Refs.~\cite{Godfrey:2016nwn,li:2021hss} reach the opposite conclusion, namely that $B^*\pi$ provides the dominant contribution. Among all theoretical models considered, we obtain the largest total width, $436~\text{MeV}$, which is significantly larger than all other predictions.

\begin{table}
\caption{Two-body strong decay widths (MeV) of the $\bar{B}_2$ state.}
\vspace{0.2cm}
\setlength{\tabcolsep}{0.01cm}
\centering
\begin{tabular*}{\textwidth}{@{}@{\extracolsep{\fill}}ccccccc}
\hline\hline
Mode & Ours & Ref. \cite{Lu:2016bbk} & Ref. \cite{Godfrey:2016nwn} & Ref. \cite{Chen:2022fye} & Ref. \cite{Asghar:2018tha} & Ref. \cite{li:2021hss} \\
\hline
{\phantom{\Large 1}}\raisebox{+.2cm}{\phantom{\Large j}}
$B^{*-}\pi^+$ & 32.4 & \multirow{2}{*}{60.57} & \multirow{2}{*}{20.1} & \multirow{2}{*}{20.2} & \multirow{2}{*}{33.64} & \multirow{2}{*}{92} \\
{\phantom{\Large 1}}\raisebox{+.2cm}{\phantom{\Large j}}
$\bar{B}^{*0}\pi^0$ & 16.1 & & & & & \\
{\phantom{\Large 1}}\raisebox{+.2cm}{\phantom{\Large j}}
$B^-\rho^+$ & 9.17 & \multirow{2}{*}{--} & \multirow{2}{*}{1.57} & \multirow{2}{*}{--} & \multirow{2}{*}{--} & \multirow{2}{*}{--} \\
{\phantom{\Large 1}}\raisebox{+.2cm}{\phantom{\Large j}}
$\bar{B}^0\rho^0$ & 4.45 & & & & & \\
{\phantom{\Large 1}}\raisebox{+.2cm}{\phantom{\Large j}}
$\bar{B}_s^{*0}\bar{K}^0$ & 7.50 & 20.75 & 0.237 & -- & 5.2 & -- \\
{\phantom{\Large 1}}\raisebox{+.2cm}{\phantom{\Large j}}
$\bar{B}^{*0}\eta$ & 2.09 & 15.58 & 0.316 & 0.1 & 0.74 & 6.7 \\
{\phantom{\Large 1}}\raisebox{+.2cm}{\phantom{\Large j}}
${B}_0^{*-}(1^3P_0)\pi^+$ & 1.90 & \multirow{2}{*}{81.79} & \multirow{2}{*}{0.0415} & \multirow{2}{*}{0.2} & \multirow{2}{*}{--} & \multirow{2}{*}{0.002} \\
{\phantom{\Large 1}}\raisebox{+.2cm}{\phantom{\Large j}}
$\bar{B}_0^{*0}(1^3P_0)\pi^0$ & 0.98 & & & & & \\
{\phantom{\Large 1}}\raisebox{+.2cm}{\phantom{\Large j}}
$B_1^-(1P_1)\pi^+$ & \multirow{2}{*}{0.58} & \multirow{2}{*}{0.05} & \multirow{2}{*}{0.333} & \multirow{2}{*}{0.4} & \multirow{2}{*}{--} & \multirow{2}{*}{0.4} \\
{\phantom{\Large 1}}\raisebox{+.2cm}{\phantom{\Large j}}
$\bar{B}_1^0(1P_1)\pi^0$ & & & & & & \\
{\phantom{\Large 1}}\raisebox{+.2cm}{\phantom{\Large j}}
$B_1^-(1P'_1)\pi^+$ & \multirow{2}{*}{0.17} & \multirow{2}{*}{--} & \multirow{2}{*}{0.03} & \multirow{2}{*}{--} & \multirow{2}{*}{--} & \multirow{2}{*}{--} \\
{\phantom{\Large 1}}\raisebox{+.2cm}{\phantom{\Large j}}
$\bar{B}_1^0(1P'_1)\pi^0$ & & & & & & \\
{\phantom{\Large 1}}\raisebox{+.2cm}{\phantom{\Large j}}
$B_2^{*-}(5747)\pi^+$ & 5.45 & \multirow{2}{*}{0.01} & \multirow{2}{*}{0.115} & \multirow{2}{*}{0.1} & \multirow{2}{*}{--} & \multirow{2}{*}{64} \\
{\phantom{\Large 1}}\raisebox{+.2cm}{\phantom{\Large j}}
$\bar{B}_2^{*0}(5747)\pi^0$ & 2.71 & & & & & \\
\hline
{\phantom{\Large 1}}\raisebox{+.2cm}{\phantom{\Large j}}
$\Gamma_{total}$ & 83.4 & 250.07 & 22.74 & 22 & 34.38 & 168.34 \\
\hline\hline
\end{tabular*}
\label{B2}
\end{table}

\begin{table}
\caption{Two-body strong decay widths (MeV) of the $\bar{B'}_2$ state.}
\vspace{0.2cm}
\setlength{\tabcolsep}{0.01cm}
\centering
\begin{tabular*}{\textwidth}{@{}@{\extracolsep{\fill}}ccccccc}
\hline\hline
Mode & Ours & Ref. \cite{Lu:2016bbk} & Ref. \cite{Godfrey:2016nwn} & Ref. \cite{Chen:2022fye} & Ref. \cite{Asghar:2018tha} & Ref. \cite{li:2021hss} \\
\hline
{\phantom{\Large 1}}\raisebox{+.2cm}{\phantom{\Large j}}
$B^{*-}\pi^+$ & 79.3 & \multirow{2}{*}{64.08} & \multirow{2}{*}{96.3} & \multirow{2}{*}{64.7} & \multirow{2}{*}{36.19} & \multirow{2}{*}{66} \\
{\phantom{\Large 1}}\raisebox{+.2cm}{\phantom{\Large j}}
$\bar{B}^{*0}\pi^0$ & 39.4 & & & & & \\
{\phantom{\Large 1}}\raisebox{+.2cm}{\phantom{\Large j}}
$B^-\rho^+$ & 0.65 & \multirow{2}{*}{48.37} & \multirow{2}{*}{0.64} & \multirow{2}{*}{4.1} & \multirow{2}{*}{--} & \multirow{2}{*}{--} \\
{\phantom{\Large 1}}\raisebox{+.2cm}{\phantom{\Large j}}
$\bar{B}^0\rho^0$ & 0.32 & & & & & \\
{\phantom{\Large 1}}\raisebox{+.2cm}{\phantom{\Large j}}
$\bar{B}_s^{*0}\bar{K}^0$ & 48.2 & 5.04 & 26.8 & 10.5 & 16.43 & 1.4 \\
{\phantom{\Large 1}}\raisebox{+.2cm}{\phantom{\Large j}}
$\bar{B}^{*0}\eta$ & 17.4 & 4.64 & 14.1 & 13.7 & 11.88 & 1.7 \\
{\phantom{\Large 1}}\raisebox{+.2cm}{\phantom{\Large j}}
$B_0^{*-}(1^3P_0)\pi^+$ & 0.88 & \multirow{2}{*}{3.51} & \multirow{2}{*}{--} & \multirow{2}{*}{0.3} & \multirow{2}{*}{--} & \multirow{2}{*}{3.5} \\
{\phantom{\Large 1}}\raisebox{+.2cm}{\phantom{\Large j}}
$\bar{B}_0^{*0}(1^3P_0)\pi^0$ & 0.44 & & & & & \\
{\phantom{\Large 1}}\raisebox{+.2cm}{\phantom{\Large j}}
$B_1^-(1P_1)\pi^+$ & \multirow{2}{*}{0.51} & \multirow{2}{*}{0.27} & \multirow{2}{*}{0.757} & \multirow{2}{*}{1.2} & \multirow{2}{*}{--} & \multirow{2}{*}{0.4} \\
{\phantom{\Large 1}}\raisebox{+.2cm}{\phantom{\Large j}}
$\bar{B}_1^0(1P_1)\pi^0$ & & & & & & \\
{\phantom{\Large 1}}\raisebox{+.2cm}{\phantom{\Large j}}
$B_1^-(1P'_1)\pi^+$ & \multirow{2}{*}{0.25} & \multirow{2}{*}{8.31} & \multirow{2}{*}{--} & \multirow{2}{*}{0.3} & \multirow{2}{*}{--} & \multirow{2}{*}{10} \\
{\phantom{\Large 1}}\raisebox{+.2cm}{\phantom{\Large j}}
$\bar{B}_1^0(1P'_1)\pi^0$ & & & & & & \\
{\phantom{\Large 1}}\raisebox{+.2cm}{\phantom{\Large j}}
$B_2^{*-}(5747)\pi^+$ & 165 & \multirow{2}{*}{0.20} & \multirow{2}{*}{73.8} & \multirow{2}{*}{188.3} & \multirow{2}{*}{--} & \multirow{2}{*}{6.2} \\
{\phantom{\Large 1}}\raisebox{+.2cm}{\phantom{\Large j}}
$\bar{B}_2^{*0}(5747)\pi^0$ & 83.7 & & & & & \\
\hline
{\phantom{\Large 1}}\raisebox{+.2cm}{\phantom{\Large j}}
$\Gamma_{total}$ & 436 & 190.76 & 213.00 & 283.1 & 64.5 & 89.20 \\
\hline\hline
\end{tabular*}
\label{B2'}
\end{table}

\subsection{Strong decays of $\bar{B}_{s2}$ and $\bar{B'}_{s2}$}
Similar to the case of $D_{s2}$ and $D'_{s2}$, since the light quark is an $s$ quark, the final-state light mesons containing an $s$ quark are not as light as pions. This leads to a very small number of OZI-allowed strong decay channels for $\bar{B}_{s2}$ and $\bar{B'}_{s2}$, with only two decay channels for each state. We therefore present the strong decay results for $\bar{B}_{s2}$ and $\bar{B'}_{s2}$ in a single table, namely Table~\ref{Bs2}. For comparison, some other theoretical results are also listed in the same table. As before, we label the lower-mass state as $\bar{B}_{s2}$ and the higher-mass state as $\bar{B'}_{s2}$. As seen from Table~\ref{Bs2}, our results together with those from Refs.~\cite{Godfrey:2016nwn,Asghar:2018tha} indicate that the lower-mass state is narrow while the higher-mass state is broad. In contrast, Refs.~\cite{li:2021hss,Lu:2016bbk,Hao:2022ibj} reach the opposite conclusion, namely that the lower-mass state is broad and the higher-mass state is narrow.

\begin{table}[htbp]
\centering
\caption{Two-body strong decay widths (MeV) of the $\bar{B}_{s2}$ and $\bar{B'}_{s2}$ states.}
\vspace{0.2cm}
\setlength{\tabcolsep}{0.2cm}
\begin{tabular*}{\textwidth}{@{}@{\extracolsep{\fill}}ccccccc@{}}
\hline\hline
Mode & ours & Ref. \cite{Godfrey:2016nwn} & Ref. \cite{Asghar:2018tha} & Ref. \cite{li:2021hss} & Ref. \cite{Lu:2016bbk} & Ref. \cite{Hao:2022ibj} \\
\hline
\multicolumn{7}{l}{$B_{s2}$} \\
\phantom{\Large{1}}\raisebox{+.2cm}{\phantom{\Large{j}}}
$B^*K$ & $41.4$ & $15.9$ & $17.89$ & $85.0$ & $180.5$ & $75$ \\
\phantom{\Large{1}}\raisebox{+.2cm}{\phantom{\Large{j}}}
$B_s^*\eta$ & $3.55$ & $0.387$ & $0.66$ & $4.2$ & $18.15$ & $9$ \\
\phantom{\Large{1}}\raisebox{+.2cm}{\phantom{\Large{j}}}
$\Gamma_{total}$ & $44.9$ & $16.4$ & $18.6$ & $90.0$ & $198.64$ & $84$ \\
\hline
\multicolumn{7}{l}{$B'_{s2}$} \\
\phantom{\Large{1}}\raisebox{+.2cm}{\phantom{\Large{j}}}
$B^*K$ & $147$ & $171$ & $108.4$ & $22.0$ & $76.41$ & $34$ \\
\phantom{\Large{1}}\raisebox{+.2cm}{\phantom{\Large{j}}}
$B_s^*\eta$ & $15.1$ & $22.3$ & $18.13$ & $0.5$ & $2.95$ & $1$ \\
\phantom{\Large{1}}\raisebox{+.2cm}{\phantom{\Large{j}}}
$\Gamma_{total}$ & $162$ & $194$ & $126.5$ & $23$ & $79.37$ & $35$ \\
\hline\hline
\end{tabular*}
\label{Bs2}
\end{table}

For the narrow state, all theoretical results show that $B^* K$ is the overwhelmingly dominant decay channel, while the $B^*_s \eta$ channel contributes very little. For instance, our results give a branching ratio of $92\%$ for the $B^* K$ decay, whereas other theoretical predictions are even larger, ranging between $96\%$ and $97\%$. We predict a total width of $44.9~\text{MeV}$ for the narrow state, while other theoretical values lie in the range of $16.4\sim79.37~\text{MeV}$. 

For the broad state, all theoretical results likewise indicate that $B^* K$ is also the dominant decay channel. For example, the branching ratio for $B^* K$ in our results is $91\%$, and other theoretical values range from $86\%$ to $97\%$, which are relatively close to each other. Regarding its total width, theoretical predictions vary from $84$ to $198.64~\text{MeV}$. Our result of $162~\text{MeV}$ is in relatively good agreement with the values of $194~\text{MeV}$ from Ref.~\cite{Godfrey:2016nwn} and $198.64~\text{MeV}$ from Ref.~\cite{Lu:2016bbk}.

\subsection{Partial-wave contributions and relativistic effects}

In presenting the explicit representations, we have pointed out that each wave function contains both non-relativistic terms and relativistic correction terms, which are classified into different partial waves. In this part of the article, we  choose several  channels to examine the role of these partial-wave components and  address the relativistic corrections. Because light mesons are not described by wave functions in the present study, the discussion below is restricted to the wave functions of the heavy mesons in the initial and final states. 

We first examine the decays $D_2(2740)\to D^{*+}\pi^-$ and $\bar{B}_2\to B^{*-}\pi^+$, which represent the largest branching-fraction channels of $D_2(2740)$ and $\bar{B}_2$, respectively. The partial-wave decomposition of these decays is presented in Tables~\ref{1-d2} and in Tables~\ref{1-b2}. As previously discussed, the wave functions of the $2^-$ states $D_2$ and $\bar{B}_2$ are composed of non-relativistic $D$-waves supplemented by relativistic $F$-waves; whereas the final $1^-$ mesons ($D^{*+}$ and $B^{*-}$) possess wave functions of the $S+P+D$ type,  in which the $S$-wave is non-relativistic while the $P+D$ parts are relativistic corrections. In the tables, the rows refer to the partial waves of the initial $2^-$ meson and the columns to those of the final $1^-$ meson, the latter being marked with primes. 

\begin{table}[h]
\centering
\caption{Partial-wave contributions to the decay width for the process $D_2(2740)\to D^{*+}\pi^-$, in units of MeV.}
\begin{tabular}{lcccc}
\hline
$1^-$ &~ Whole &~ $S'$-wave ($C_5,C_6$) &~ $P'$-wave ($C_1,C_2,C_7,C_8$) &~ $D'$-wave ($C_3,C_4$) \\
\hline
$2^-$ & & & & \\
Whole & 23.8 & 13.8 & 1.90 & $\sim 0$ \\
$D$-wave ($A_1,A_2,A_5,A_6$) & 32.3 & 17.4 & 8.76 & $\sim 0$ \\
$F$-wave ($A_3,A_4,A_7,A_8$) & 4.32  & 7.82  & 2.53 & $\sim 0$ \\
\hline
\end{tabular}
\label{1-d2}
\end{table}
\begin{table}[h]
\centering
\caption{Partial-wave contributions to the decay width for the process $\bar{B}_2\to B^{*-}\pi^+$, in units of MeV.}
\begin{tabular}{lcccc}
\hline
$1^-$ &~ Whole &~ $S'$-wave ($C_5,C_6$) &~ $P'$-wave ($C_1,C_2,C_7,C_8$) &~ $D'$-wave ($C_3,C_4$) \\
\hline
$2^-$ & & & & \\
Whole & 32.4 & 18.4 & 2.40 & $\sim 0$ \\
$D$-wave ($A_1,A_2,A_5,A_6$) & 59.8 & 30.9 & 11.8 & $\sim 0$ \\
$F$-wave ($A_3,A_4,A_7,A_8$) & 8.87  & 9.91  & 3.54 & $\sim 0$ \\
\hline
\end{tabular}
\label{1-b2}
\end{table}
The ``Whole'' row (column) keeps all partial waves of that meson; the ``$D$-wave'' row keeps only the $D$-wave ($A_{1,2,5,6}$) and drops the relativistic $F$-wave ($A_{3,4,7,8}$) of $2^-$ meson, whereas the ``$F$-wave'' row does the converse; an ``$S'$-wave'' column retains only the $S'$-wave ($C_{5,6}$) of $1^-$ meson and discards its $P'$- and $D'$-waves  ($C_{1,2,3,4,7,8}$), with analogous meanings for the other labels. As can be seen from the Table~\ref{1-d2}, we present only a small subset of typical contributions, for example, 
the sole non-relativistic contribution $(D\!\times\!S')^2$ $\to 17.4$ MeV (hereafter abbreviated as $D\!\times\!S'$); for the most part, all cross-term contributions, such as $(D\!\times\!S')\!\times\!(D\!\times\!P')$, $(D\!\times\!S')\!\times\!(F\!\times\!P')$, and so on, are not listed.  Values smaller than 0.02~MeV are denoted by $\sim0$.

It can be seen from Table~\ref{1-d2}, that the total width of $D_2(2740)\to D^{*+}\pi^-$ is $\Gamma=23.8$~MeV; this is the entry in which both the row and the column are labelled ``Whole'', i.e. no partial wave of either meson is discarded. In the non-relativistic limit, where the $2^-$ state contributes only through its $D$-wave and the $1^-$ state only through its $S'$-wave (the entry marked $D\!\times\! S'$), one obtains the non-relativistic decay width $\Gamma_{\rm non}=17.4$~MeV. We define the relativistic effect as $R\equiv(\Gamma-\Gamma_{\rm non})/\Gamma$, which gives $27\%$ for $D_2(2740)\to D^{*+}\pi^-$ and $4.6\%$ for $\bar{B}_2\to B^{*-}\pi^+$ (Table~\ref{1-b2}).

We stress that the smallness of the relativistic effect especially in channel $\bar{B}_2\to B^{*-}\pi^+$ is caused by a cancellation rather than by genuinely small relativistic entries: in Table~\ref{1-d2} the $D\!\times\! P'$ and $F\!\times\! S'$ contributions are 8.76 MeV and 7.82~MeV, i.e. 37\% and 33\% of total $\Gamma=23.8$ MeV, while for $\bar{B}_2\to B^{*-}\pi^+$ in Table~\ref{1-b2}, the corresponding entries are 11.8 MeV and 9.91~MeV, amounting $36\%$ and $31\%$ of total $\Gamma=32.4$~MeV, all of which are sizable before cancellation.

\begin{table}[h]
\centering
\caption{Partial-wave contributions to the decay width for the process $D'_2 \to D_2^{*+}(2460)\pi^{-}$, in units of MeV. }
\begin{tabular}{lcccc}
\hline
$2^+$ &~ Whole &~ $P'$-wave ($F_5,F_6$) &~ $D'$-wave ($F_1,F_2,F_7,F_8$) &~ $F'$-wave ($F_3,F_4$) \\
\hline
$2^-$ & & & & \\
Whole & 109 & 15.7 & 35.2 & 0.08 \\
$D$-wave ($A_1,A_2,A_5,A_6$) & 21.3 & 9.90 & 44.9 & $\sim 0$ \\
$F$-wave ($A_3,A_4,A_7,A_8$) & 38.2  & 39.2  & 0.14 & 0.07 \\
\hline
\end{tabular}
\label{2+d2}
\end{table}
\begin{table}[h]
\centering
\caption{Partial-wave contributions to the decay width for the process $\bar{B'}_2 \to B_2^{*-}(5747)\pi^{+}$, in units of MeV.}
\begin{tabular}{lcccc}
\hline
$2^+$ &~ Whole &~ $P'$-wave ($F_5,F_6$) &~ $D'$-wave ($F_1,F_2,F_7,F_8$) &~ $F'$-wave ($F_3,F_4$) \\
\hline
$2^-$ & & & & \\
Whole & 165 & 28.0 & 65.9 & $\sim 0$ \\
$D$-wave ($A_1,A_2,A_5,A_6$) & 35.6 & 33.6 & 68.4 & $\sim 0$ \\
$F$-wave ($A_3,A_4,A_7,A_8$) & 73.1 & 73.7  & 0.236 & $\sim 0$ \\
\hline
\end{tabular}
\label{2+b2}
\end{table}
The entries in the $D'$ column are uniformly tiny ($\lesssim0.02$~MeV), so that the $D'$-wave component of the vector meson is negligible; equivalently, the $C_3$ and $C_4$ terms of the vector-meson wave function in Eq.~(\ref{positive energy wave 1-}) can be disregarded. This statement holds for all the $1^-$ final states only, for the $2^+$ mesons the $D'$-wave entries are large ($F_1$, $F_2$, $F_7$, and $F_8$ terms), as shown in Tables~\ref{2+d2} and \ref{2+b2}.

Tables~\ref{2+d2} and \ref{2+b2} list the partial-wave contributions for $D'_2 \to D_2^{*+}(2460)\pi^{-}$ and $\bar{B'}_2 \to B_2^{*-}(5747)\pi^{+}$, respectively. In both cases the initial state is the heavier member of a $2^-$ doublet and the final state is a $2^+$ heavy meson. In Table~\ref{2+d2}, the non-relativistic contribution to $D'_2 \to D_2^{*+}(2460)\pi^{-}$ arises from the $D\!\times\! P'$ entry, $\Gamma_{\rm non}=9.90$~MeV, whereas the total width is $\Gamma=109$~MeV, so that the relativistic effect is $R=91\%$; for $\bar{B'}_2 \to B_2^{*-}(5747)\pi^{+}$ (Table~\ref{2+b2}) one finds $\Gamma_{\rm non}=33.6$~MeV, $\Gamma=165$~MeV and a relativistic effect of $R=80\%$. The large relativistic corrections are also visible in the individual entries: the $D\!\times\! D'$ and $F\!\times\! P'$ contributions, 44.9 MeV and 39.2~MeV for $D'_2 \to D_2^{*+}(2460)\pi^{-}$ and 68.4 MeV and 73.7~MeV for $\bar{B'}_2 \to B_2^{*-}(5747)\pi^{+}$, are much larger than the non-relativistic $D\!\times\! P'$ contributions 9.90 MeV and 33.6 MeV, respectively. Similar to the $D'$-wave component in $1^-$ meson, the $F'$-wave component ($F_3$ and $F_4$ terms in Tables~\ref{2+d2} and \ref{2+b2}) of the $2^+$ meson is also negligible. 

\begin{table}[h]
\centering
\caption{Partial-wave contributions to the decay width for the process $D_{s2} \to D^{*0} K^+$, in units of MeV.}
\begin{tabular}{lcccc}
\hline
$1^-$ &~ Whole &~ $S'$-wave ($C_5,C_6$) &~ $P'$-wave ($C_1,C_2,C_7,C_8$) &~ $D'$-wave ($C_3,C_4$) \\
\hline
$2^-$ & & & & \\
Whole & 25.5 & 11.1 & 4.30 & $\sim 0$ \\
$D$-wave ($A_1,A_2,A_5,A_6$) & 28.7 & 14.3 & 10.5 & $\sim 0$ \\
$F$-wave ($A_3,A_4,A_7,A_8$) & 3.77  & 6.39  & 1.43 & $\sim 0$ \\
\hline
\end{tabular}
\label{1-ds2+}
\end{table}

\begin{table}[h]
\centering
\caption{Partial-wave contributions to the decay width for the process $\bar{B}_{s2} \to B^{*-} K^+$, in units of MeV. }
\begin{tabular}{lcccc}
\hline
$1^-$ &~ Whole &~ $S'$-wave ($C_5,C_6$) &~ $P'$-wave ($C_1,C_2,C_7,C_8$) &~ $D'$-wave ($C_3,C_4$) \\
\hline
$2^-$ & & & & \\
Whole & 20.9 & 9.12 & 3.05 & $\sim 0$ \\
$D$-wave ($A_1,A_2,A_5,A_6$) & 30.2 & 14.4 & 8.90 & $\sim 0$ \\
$F$-wave ($A_3,A_4,A_7,A_8$) & 3.42  & 5.22  & 1.53 & $\sim 0$ \\
\hline
\end{tabular}
\label{1-bs2+}
\end{table}
To us, the large relativistic effects $R=91\%$ and $R=80\%$ exhibited in $D'_2 \to D_2^{*+}(2460)\pi^{-}$ and $\bar{B'}_2 \to B_2^{*-}(5747)\pi^{+}$ are not surprising: we have previously found that for heavy mesons, the higher the excitation level, the larger the relativistic effect~\cite{Geng:2018qrl,Wang:2020zbr,Li:2024dyi}. Both the initial ($2^-$) and final ($2^+$) heavy mesons in these two processes are highly excited states, so large relativistic effects are to be expected. On the other hand, the relatively small relativistic effects obtained earlier, $27\%$ for $D_2(2740)\to D^{*+}\pi^-$ and $4.6\%$ for $\bar{B}_2\to B^{*-}\pi^+$, particularly the latter, are much smaller than  our expectation. As we have clarified earlier, the possible reason for the small relativistic effect is that there are many relativistic correction terms, and cancellations among them lead to a very small 
final result.

\begin{table}[h]
\centering
\caption{Partial-wave contributions to the decay width for the process $D'_{s2} \to D^{*0} K^+$, in units of MeV.}
\begin{tabular}{lcccc}
\hline
$1^-$ &~ Whole &~ $S'$-wave ($C_5,C_6$) &~ $P'$-wave ($C_1,C_2,C_7,C_8$) &~ $D'$-wave ($C_3,C_4$) \\
\hline
$2^-$ & & & & \\
Whole & 66.2 & 8.79 & 23.5 & 0.19 \\
$D$-wave ($A_1,A_2,A_5,A_6$) & 17.7 & 19.1 & 56.4 & $\sim 0$ \\
$F$-wave ($A_3,A_4,A_7,A_8$) & 20.8  & 45.6  & 7.10 & 0.17 \\
\hline
\end{tabular}
\label{1-ds2}
\end{table}
\begin{table}[h]
\centering
\caption{Partial-wave contributions to the decay width for the process $\bar{B}'_{s2} \to B^{*-} K^+$, in units of MeV. }
\begin{tabular}{lcccc}
\hline
$1^-$ &~ Whole &~ $S'$-wave ($C_5,C_6$) &~ $P'$-wave ($C_1,C_2,C_7,C_8$) &~ $D'$-wave ($C_3,C_4$) \\
\hline
$2^-$ & & & & \\
Whole & 73.6 & 12.7 & 24.7 & $\sim 0$ \\
$D$-wave ($A_1,A_2,A_5,A_6$) & 21.5 & 19.8 & 72.1 & $\sim 0$ \\
$F$-wave ($A_3,A_4,A_7,A_8$) & 18.4  & 59.1  & 12.4 & $\sim 0$ \\
\hline
\end{tabular}
\label{1-bs2}
\end{table}
In the case of $D_{s2}$ and $\bar{B}_{s2}$, the processes with the largest
branching fractions are $D_{s2}\to D^{*0}K^+$ and
$\bar{B}_{s2}\to B^{*-}K^+$, whose partial-wave contributions are given in
Tables~\ref{1-ds2+} and~\ref{1-bs2+}. The corresponding relativistic effects
are $44\%$ and $31\%$, respectively. Their higher partners behave quite
differently: for $D'_{s2}\to D^{*0}K^+$ (Table~\ref{1-ds2}) and
$\bar{B}'_{s2}\to B^{*-}K^+$ (Table~\ref{1-bs2}), the corresponding
relativistic effects are $71\%$ and $73\%$, respectively, both quite large.

Comparing the decays of the lower- and higher-mass members of each doublet,
we find the same pattern for both the $2^{-}\to1^{-}$ and the
$2^{-}\to2^{+}$ transitions (the 
initial and final states in both cases being the heavy mesons): the relativistic corrections are small for the lighter member and large for the heavier one. Taking the $\bar{B}_{s2}$ and
$\bar{B}'_{s2}$ decays as an example, the relativistic effects amount to
$31\%$ and $73\%$, respectively. A difference this large, together with the
fact that every doublet exhibits the same pattern, indicates that it is not
kinematic but dynamical in origin. Eq.~\eqref{nrmixing} gives the
$D$-wave mixing formula of the general theory, whereas in our model there is
not only the nonrelativistic $D$-wave but also a relativistic $F$-wave;
Eq.~\eqref{nrmixing} then implies that the lighter and heavier states
contain different amounts of $D$- and $F$-wave components, which is the underlying reason for the large spread in relativistic effects.  This conclusion is supported both by the mixing angles, which are
very close within each doublet, and by the regular pattern of the
partial-wave contributions to the decays. For instance, in the
$2^{-}\to1^{-}$ transitions the main relativistic contributions of the
lighter state, $D\times P'$ and $F\times S'$, are both smaller than its
nonrelativistic $D\times S'$, whereas for the heavier state both
$D\times P'$ and $F\times S'$ are markedly larger than $D\times S'$. The
same pattern holds for the $2^{-}\to2^{+}$ transitions and will not be detailed here.

\section{Summary}\label{sec:summary}

In this work we have carried out a relativistic study of the $J^P=2^-$
heavy-light mesons, including their mass spectra, the
$^1D_2$--$^3D_2$ mixing angles, and their OZI-allowed two-body strong
decays. The whole calculation is based on the instantaneous
Bethe--Salpeter (Salpeter) equation, which is solved numerically for the
radial wave functions; the wave-function representations are written in
terms of the $J^P$ quantum numbers instead of the conventional
non-relativistic spectroscopic notation ${}^{2S+1}L_J$. The hadronic matrix
element is written as an overlap integral
of the positive-energy wave functions of the initial and final heavy
mesons.

For the mass spectrum, the lighter member of each $2^-$ doublet is
taken as input, and the heavier partner together with the mixing angle
is predicted. We obtain $M_{D'_2}=2803$ MeV, $M_{D'_{s2}}=2936$ MeV,
$M_{\bar{B}'_2}=6146$ MeV, and $M_{\bar{B}'_{s2}}=6203$ MeV, corresponding to mass
splittings $\Delta M=56$, $36$, $51$, and $34$ MeV, respectively. The mixing angles extracted from the
normalization condition of the $2^-$ wave function are
$\theta=59.2^\circ$, $60.5^\circ$, $56.1^\circ$, and $56.3^\circ$
for the $c\bar q$, $c\bar s$, $b\bar q$, and $b\bar s$ systems,
respectively; their signs are fixed by comparing our full wave
function with the individual $^1D_2$ and $^3D_2$ wave functions. We stress that a change of the sign of the mixing
angle is equivalent to interchanging the lighter and heavier mixed
states, or to replacing the $Q\bar q$ configuration by $\bar Qq$.

For the strong decays, the predicted total widths are
$\Gamma_{D_2}=99.2$ MeV, $\Gamma_{D'_2}=291$ MeV,
$\Gamma_{D_{s2}}=98.0$ MeV, $\Gamma_{D'_{s2}}=149$ MeV,
$\Gamma_{\bar B_2}=83.4$ MeV, $\Gamma_{\bar B'_2}=436$ MeV,
$\Gamma_{\bar B_{s2}}=44.9$ MeV, and $\Gamma_{\bar B'_{s2}}=162$ MeV.
For $D_2(2740)$ the dominant channels are $D^*\pi$ and $D\rho$, the resulting total width is consistent with the experimental value
$89\pm19$ MeV. Its heavier
partner $D'_2$ is dominated by $D_2^*(2460)\pi$ and $D^*\pi$. For the charm-strange doublet, the width of $D_{s2}$ is shared by $D^*K$ and $DK^*$, while
$D'_{s2}$ decays predominantly into $D^*K$. In the bottom sector, $\bar B_2$ decays
mainly through $B^*\pi$ and $B\rho$, while $\bar B'_2$ is dominated by
$B_2^*(5747)\pi$ and $B^*\pi$. In the $b\bar s$ sector, $B^*K$ overwhelmingly
dominates both $\bar B_{s2}$ and $\bar B'_{s2}$. Our results show that the lower-mass states are narrow, while the higher-mass states are broad.

Finally, we have examined the partial-wave composition of several
representative channels in order to quantify the relativistic
corrections. And the results confirm our earlier observation that
the higher the orbital excitation of the heavy mesons involved, the more
important the relativistic treatment becomes, and demonstrate that a
non-relativistic framework is inadequate for the $2^-$ heavy-light mesons.

\vspace{0.7cm} {\bf Acknowledgments}

This work was supported by the National Natural Science Foundation of China (NSFC) under the Grants No. 12575097 and No. 12375085. Q. Li was supported by the National Key R\&D Program of China\,(2022YFA1604803) and the Natural Science Basic Research Program of Shaanxi\,(No.\,2025JC-YBMS-020). T. Wang was also supported by the Fundamental Research Funds for the Central Universities (2023FRFK06009).

\appendix
\section{Coefficients of the positive-energy wave functions}
The positive-energy wave functions of the $2^-$ mesons are presented in Eq.~(\ref{wave++}), in which the coefficients $A_i$ depend on the radial wave functions, and their explicit representations read:
\begin{equation}
\label{2-}
\begin{aligned}
A_1 &= \frac{1}{2}\left[ f_1 + \frac{\omega_1 + \omega_2}{m_1 + m_2} f_2 \right], &
A_3 &= -\frac{M (\omega_1 - \omega_2)}{m_1 \omega_2 + m_2 \omega_1} A_1, \\[6pt]
A_2 &= \frac{1}{2}\left[ f_2 + \frac{m_1 + m_2}{\omega_1 + \omega_2} f_1 \right], &
A_4 &= -\frac{M (m_1 + m_2)}{m_1 \omega_2 + m_2 \omega_1} A_1.\\
A_5 &= \frac{1}{2}\left[ v_1 - \frac{\omega_1 + \omega_2}{m_1 + m_2} v_2 \right], &
A_7 &= +\frac{M (\omega_1 - \omega_2)}{m_1 \omega_2 + m_2 \omega_1} A_5, \\[6pt]
A_6 &= \frac{1}{2}\left[ v_2 - \frac{m_1 + m_2}{\omega_1 + \omega_2} v_1 \right], &
A_8 &= -\frac{M (m_1 + m_2)}{m_1 \omega_2 + m_2 \omega_1} A_5.
\end{aligned}
\end{equation}
For the positive-energy wave functions of the pseudoscalar $0^-$ mesons (Eq.~(\ref{positive energy wave 0-})), the coefficients are expressed as 
\begin{equation}
\label{0-}
\begin{aligned}
B_1 &= \frac{M}{2}\left(\frac{h_1(\omega_1+\omega_2)}{m_1+m_2}+h_2\right), &\quad
B_3 &= -\frac{B_1 M (\omega_1-\omega_2)}{m_1 \omega_2 + m_2 \omega_1}, \\
B_2 &= \frac{M}{2}\left(\frac{h_2(\omega_1+\omega_2)}{m_1+m_2}+h_1\right), &
B_4 &= -\frac{B_1 M (m_1+m_2)}{m_1 \omega_2 + m_2 \omega_1},
\end{aligned}
\end{equation}
where $h_1$ and $h_2$ denote the radial wave functions obtained from solving the Salpeter equation. Eq.~(\ref{positive energy wave 1-}) presents the positive-energy wave functions of the $1^-$ vector mesons, the coefficient expressions of which are given by 
\begin{equation}
\label{1-}
\begin{aligned}
C_1 &= \frac{1}{2 M (m_1 \omega_2 + m_2 \omega_1)}
\Big( - (\omega_1 + \omega_2) \boldsymbol{q}^2 g_3 - (m_1 + m_2) \boldsymbol{q}^2 g_4
+ 2 M^2 \omega_2 g_5 - 2 M^2 m_2 g_6 \Big),\\
C_2 &= \frac{1}{2 M (m_1 \omega_2 + m_2 \omega_1)}
\Big( - (m_1 - m_2) \boldsymbol{q}^2 g_3 - (\omega_1 - \omega_2) \boldsymbol{q}^2 g_4
- 2 M^2 m_2 g_5 + 2 M^2 \omega_2 g_6 \Big),\\
C_3 &= \frac{1}{2} \Big( g_3 + \frac{m_1 + m_2}{\omega_1 + \omega_2} g_4
- \frac{2 M^2}{m_1 \omega_2 + m_2 \omega_1} g_6 \Big),\\
C_4 &= \frac{1}{2} \Big( \frac{\omega_1 + \omega_2}{m_1 + m_2} g_3 + g_4
- \frac{2 M^2}{m_1 \omega_2 + m_2 \omega_1} g_5 \Big),\\
C_5 &= \frac{1}{2} \Big( g_5 - \frac{\omega_1 + \omega_2}{m_1 + m_2} g_6 \Big),\\
C_6 &= \frac{1}{2} \Big( - \frac{m_1 + m_2}{\omega_1 + \omega_2} g_5 + g_6 \Big),\\
C_7 &= \frac{M}{2} \frac{\omega_1 - \omega_2}{m_1 \omega_2 + m_2 \omega_1}
\Big( g_5 - \frac{\omega_1 + \omega_2}{m_1 + m_2} g_6 \Big),\\
C_8 &= \frac{M}{2} \frac{m_1 + m_2}{m_1 \omega_2 + m_2 \omega_1}
\Big( - g_5 + \frac{\omega_1 + \omega_2}{m_1 + m_2} g_6 \Big),
\end{aligned}
\end{equation}
where $\boldsymbol{q}^2=\vec{q}^2=-q^2_{\perp}$, and $g_i$ ($i=3,4,5,6$) are radial wave functions.
The coefficient expressions for the positive-energy wave functions in Eq.~(\ref{0+w}) of the $0^+$ mesons are   
\begin{equation}
\label{0+}
\begin{aligned}
G_1 &= \frac{-\boldsymbol{q}^2(\omega_1 + \omega_2)}{2 (m_1 \omega_2 + m_2 \omega_1)}
\Big(i_1+\frac{m_1+m_2}{\omega_1+\omega_2}i_2\Big),\quad
G_3 &= \frac{1}{2}\Big(i_1+\frac{m_1+m_2}{\omega_1+\omega_2}i_2\Big),\\
G_2 &= \frac{-\boldsymbol{q}^2(m_1-m_2)}{2 (m_1 \omega_2 + m_2 \omega_1)}
\Big(i_1+\frac{m_1+m_2}{\omega_1+\omega_2}i_2\Big),\quad
G_4 &= \frac{1}{2}\Big(\frac{\omega_1+\omega_2}{m_1+m_2}i_1+i_2\Big),
\end{aligned}
\end{equation}
where $i_1$ and $i_2$ are radial wave functions.
Those of $1^+$ mesons (Eq.~(\ref{1+w})) are
\begin{equation}
\label{1+}
\begin{aligned}
E_1 &= \frac{1}{2} \Big[ k_1 + \frac{\omega_1 + \omega_2}{m_1 + m_2} k_2 \Big], &\quad
E_3 &= - \frac{M (\omega_1 - \omega_2)}{m_1 \omega_2 + m_2 \omega_1} E_1, \\
E_2 &= \frac{1}{2} \Big[ \frac{m_1 + m_2}{\omega_1 + \omega_2} k_1 + k_2 \Big], &\quad
E_4 &= - \frac{M (m_1 + m_2)}{m_1 \omega_2 + m_2 \omega_1} E_1, \\
E_5 &= \frac{1}{2} \Big[ l_1 + \frac{\omega_1 + \omega_2}{m_1 + m_2} l_2 \Big], &\quad
E_7 &= \frac{M (\omega_1 - \omega_2)}{m_1 \omega_2 + m_2 \omega_1} E_5, \\
E_6 &= - \frac{1}{2} \Big[ \frac{m_1 + m_2}{\omega_1 + \omega_2} l_1 + l_2 \Big], &\quad
E_8 &= - \frac{M (m_1 + m_2)}{m_1 \omega_2 + m_2 \omega_1} E_5,
\end{aligned}
\end{equation}
where $k_1$, $k_2$, $l_1$, and $l_2$ are radial wave functions. 
Those of the $2^+$ mesons (Eq.~(\ref{2+w})) are
\begin{equation}
\label{2+}
\begin{aligned}
F_1 &= \frac{1}{2 M (m_1 \omega_2 + m_2 \omega_1)}
\Big( - (\omega_1 + \omega_2) \boldsymbol{q}^2 \xi_3 - (m_1 + m_2) \boldsymbol{q}^2 \xi_4
+ 2 M^2 \omega_2 \xi_5 - 2 M^2 m_2 \xi_6 \Big),\\
F_2 &= \frac{1}{2 M (m_1 \omega_2 + m_2 \omega_1)}
\Big( - (m_1 - m_2) \boldsymbol{q}^2 \xi_3 - (\omega_1 - \omega_2) \boldsymbol{q}^2 \xi_4
- 2 M^2 m_2 \xi_5 + 2 M^2 \omega_2 \xi_6 \Big),\\
F_3 &= \frac{1}{2} \Big( \xi_3 + \frac{m_1 + m_2}{\omega_1 + \omega_2} \xi_4
- \frac{2 M^2}{m_1 \omega_2 + m_2 \omega_1} \xi_6 \Big),\\
F_4 &= \frac{1}{2} \Big( \frac{\omega_1 + \omega_2}{m_1 + m_2} \xi_3 + \xi_4
- \frac{2 M^2}{m_1 \omega_2 + m_2 \omega_1} \xi_5 \Big),\\
F_5 &= \frac{1}{2} \Big( \xi_5 - \frac{\omega_1 + \omega_2}{m_1 + m_2} \xi_6 \Big),\\
F_6 &= \frac{1}{2} \Big( - \frac{m_1 + m_2}{\omega_1 + \omega_2} \xi_5 + \xi_6 \Big),\\
F_7 &= \frac{M}{2} \frac{\omega_1 - \omega_2}{m_1 \omega_2 + m_2 \omega_1}
\Big( \xi_5 - \frac{\omega_1 + \omega_2}{m_1 + m_2} \xi_6 \Big),\\
F_8 &= \frac{M}{2} \frac{m_1 + m_2}{m_1 \omega_2 + m_2 \omega_1}
\Big( - \xi_5 + \frac{\omega_1 + \omega_2}{m_1 + m_2} \xi_6 \Big),
\end{aligned}
\end{equation}
where $\xi_i$ ($i=3,4,5,6$) are radial wave functions.


\begin{thebibliography}{99}

\bibitem{Chen:2022asf}
H.~X.~Chen, W.~Chen, X.~Liu, Y.~R.~Liu, and S.~L.~Zhu,
Rept. Prog. Phys. \textbf{86}, 026201 (2023).

\bibitem{Brambilla:2019esw}
N.~Brambilla, S.~Eidelman, C.~Hanhart, A.~Nefediev, C.~P.~Shen, C.~E.~Thomas, A.~Vairo, and C.~Z.~Yuan,
Phys. Rept. \textbf{873}, 1 (2020).


\bibitem{Belle:2013ewt}
V.~Bhardwaj \textit{et al.} [Belle],
Phys. Rev. Lett. \textbf{111}, no.3, 032001 (2013).

\bibitem{BESIII:2015iqd}
M.~Ablikim \textit{et al.} [BESIII],
Phys. Rev. Lett. \textbf{115}, no.1, 011803 (2015).

\bibitem{LHCb:2020fvo}
R.~Aaij \textit{et al.} [LHCb],
JHEP \textbf{08}, 123 (2020).

\bibitem{Li:2022qhg}
W.~Li, S.~Y.~Pei, T.~Wang, Y.~L.~Wang, T.~F.~Feng, and G.~L.~Wang,
Phys. Rev. D \textbf{107}, no.11, 11 (2023).


\bibitem{CLEO:2004npj}
G.~Bonvicini \textit{et al.} [CLEO],
Phys. Rev. D \textbf{70}, 032001 (2004).

\bibitem{BaBar:2010tqb}
P.~del Amo Sanchez \textit{et al.} [BaBar],
Phys. Rev. D \textbf{82}, 111102 (2010).

\bibitem{BaBar:2010zpy}
P.~del Amo Sanchez \textit{et al.} [BaBar Collaboration],
Phys. Rev. D {\textbf{82}}, 111101 (2010).

\bibitem{LHCb:2013jjb}
R.~Aaij \textit{et al.} [LHCb Collaboration],
JHEP \textbf{09}, 145 (2013).

\bibitem{LHCb:2019juy}
R.~Aaij \textit{et al.} [LHCb Collaboration],
Phys. Rev. D \textbf{101}, 032005 (2020).

\bibitem{ParticleDataGroup:2024cfk}
S.~Navas \textit{et al.} [Particle Data Group],
Phys. Rev. D \textbf{110}, 030001 (2024).

\bibitem{Zeng:1994vj}
J.~Zeng, J.~W.~Van Orden, and W.~Roberts,
Phys. Rev. D \textbf{52}, 5229 (1995).

\bibitem{DiPierro:2001dwf}
M.~Di Pierro, and E.~Eichten,
Phys. Rev. D \textbf{64}, 114004 (2001).

\bibitem{Ebert:2009ua}
D.~Ebert, R.~N.~Faustov, and V.~O.~Galkin,
Eur. Phys. J. C \textbf{66}, 197 (2010).

\bibitem{Godfrey:2015dva}
S.~Godfrey, and K.~Moats,
Phys. Rev. D \textbf{93}, 034035 (2016).

\bibitem{Ni:2021pce}
R.~H.~Ni, Q.~Li, and X.~H.~Zhong,
Phys. Rev. D \textbf{105}, 056006 (2022).

\bibitem{M:2026uqc}
S.~M, and A.~Ranjan,
Int. J. Theor. Phys. \textbf{65}, 55 (2026).

\bibitem{Godfrey:2016nwn}
S.~Godfrey, K.~Moats, and E.~S.~Swanson,
Phys. Rev. D \textbf{94}, 054025 (2016).

\bibitem{Lu:2014zua}
Q.~F.~L{\"u}, and D.~M.~Li,
Phys. Rev. D \textbf{90}, 054024 (2014).

\bibitem{Song:2015nia}
Q.~T.~Song, D.~Y.~Chen, X.~Liu, and T.~Matsuki,
Phys. Rev. D \textbf{91}, 054031 (2015).

\bibitem{Hao:2022ibj}
W.~Hao, Y.~Lu, and E.~Wang,
Eur. Phys. J. C \textbf{83}, 520 (2023).

\bibitem{Li:2018eqc}
Q.~Li, T.~Wang, Y.~Jiang, G.~L.~Wang, and C.~H.~Chang,
Phys. Rev. D \textbf{100}, no.7, 076020 (2019).

\bibitem{Salpeter:1951sz}
E.~E.~Salpeter, and H.~A.~Bethe,
Phys. Rev. \textbf{84}, 1232 (1951).

\bibitem{Salpeter:1952ib}
E.~E.~Salpeter,
Phys. Rev. \textbf{87}, 328 (1952).

\bibitem{He:2019ify}
J.~He,
Eur. Phys. J. C \textbf{79}, 393 (2019).

\bibitem{Hilger:2014nma}
T.~Hilger, C.~Popovici, M.~Gomez-Rocha, and A.~Krassnigg,
Phys. Rev. D \textbf{91}, 034013 (2015).

\bibitem{Blank:2011ha}
M.~Blank, and A.~Krassnigg,
Phys. Rev. D \textbf{84}, 096014 (2011).

\bibitem{Micu}
L. Micu, Nucl. Phys. B \textbf{10}, 521 (1969).

\bibitem{Chang:2005sd}
C.~H.~Chang, C.~S.~Kim, and G.~L.~Wang,
Phys. Lett. B \textbf{623}, 218 (2005).

\bibitem{Wang:2016enc}
T.~Wang, Z.~H.~Wang, Y.~Jiang, L.~Jiang, and G.~L.~Wang,
Eur. Phys. J. C \textbf{77}, 38 (2017).

\bibitem{Li:2017zng}
S.~C.~Li, T.~Wang, Y.~Jiang, X.~Tan, Q.~Li, G.~L.~Wang, and C.~H.~Chang,
Phys. Rev. D \textbf{97}, 054002 (2018).

\bibitem{Liu:2022kvs}
T.~T.~Liu, S.~Y.~Pei, W.~Li, M.~Han, and G.~L.~Wang,
Eur. Phys. J. C \textbf{82}, 737 (2022).

\bibitem{Wang:2022tuy}
G.~L.~ Wang, Q.~ Li, T.~ Wang, T.~F.~ Feng, X.~G.~ Wu, and C.~H.~ Chang, Eur. Phys. J. C \textbf{82}, 1027 (2022).


\bibitem{Mandelstam:1955sd}
S.~Mandelstam,
Proc. Roy. Soc. Lond. A \textbf{233}, 248 (1955).

\bibitem{Chang:2006tc}
C.~H.~Chang, J.~K.~Chen, and G.~L.~Wang,
Commun. Theor. Phys. \textbf{46}, 467 (2006).

\bibitem{Li:1994cy}
Z.~P.~Li,
Phys. Rev. D \textbf{50}, 5639 (1994).

\bibitem{Li:1997gd}
Z.~P.~Li, H.~X.~Ye, and M.~H.~Lu,
Phys. Rev. C \textbf{56}, 1099 (1997).

\bibitem{Zhao:2002id}
Q.~Zhao, J.~S.~Al-Khalili, Z.~P.~Li, and R.~L.~Workman,
Phys. Rev. C \textbf{65}, 065204 (2002).

\bibitem{Matsuki:2011xp}
T.~Matsuki, and K.~Seo,
Phys. Rev. D \textbf{85}, 014036 (2012).

\bibitem{Zhao:1998fn}
Q.~Zhao, Z.~p.~Li and C.~Bennhold,
Phys. Rev. C \textbf{58}, 2393 (1998).


\bibitem{Zhong:2009sk}
X.~H.~Zhong, and Q.~Zhao,
Phys. Rev. D \textbf{81}, 014031 (2010).

\bibitem{Tan:2022idi}
X.~Z.~Tan, T.~Wang, Y.~Jiang, Q.~Li, L.~Huo, and G.~L.~Wang,
Phys. Rev. D \textbf{105}, 094019 (2022).

\bibitem{Wang:2022cxy}
G.~L.~Wang, T.~Wang, Q.~Li, and C.~H.~Chang,
JHEP \textbf{05}, 006 (2022).

\bibitem{Ke:2023mrt}
W.~Y.~Ke, S.~Y.~Pei, T.~Wang, and G.~L.~Wang,
Phys. Rev. D \textbf{109}, 053004 (2024).

\bibitem{Tan:2019hbw}
X.~Z.~Tan, Y.~Jiang, T.~Wang, T.~Zhou, G.~Li, Z.~K.~Geng, and G.~L.~Wang,
Eur. Phys. J. C \textbf{79}, 805 (2019).

\bibitem{Liu:2016efm}
J.~B.~Liu, and C.~D.~Lu,
Eur. Phys. J. C \textbf{77}, no.5, 312 (2017)

\bibitem{li:2021hss}
Q.~Li, R.~H.~Ni, and X.~H.~Zhong,
Phys. Rev. D \textbf{103}, 116010 (2021).

\bibitem{Li:2016efw}
Q.~Li, T.~Wang, Y.~Jiang, H.~Yuan, T.~Zhou, and G.~L.~Wang,
Eur. Phys. J. C \textbf{77}, 12 (2017).

\bibitem{Yu:2014dda}
G.~L.~Yu, Z.~G.~Wang, and Z.~Y.~Li,
Chin. Phys. C \textbf{39}, 063101 (2015).



\bibitem{Hao:2024ptu}
W.~Hao, M.~A.~Sultan, and E.~Wang,
Phys. Rev. D \textbf{111}, 094024 (2025).

\bibitem{Yang:2023tvc}
J.~J.~Yang, W.~Hao, X.~Wang, D.~M.~Li, Y.~X.~Li, and E.~Wang,
Eur. Phys. J. C \textbf{83}, 1098 (2023).

\bibitem{Hao:2022vwt}
W.~Hao, Y.~Lu, and B.~S.~Zou,
Phys. Rev. D \textbf{106}, 074014 (2022).

\bibitem{Chen:2022fye}
B.~Chen, S.~Q.~Luo, K.~W.~Wei, and X.~Liu,
Phys. Rev. D \textbf{105}, 074014 (2022).

\bibitem{Asghar:2018tha}
I.~Asghar, B.~Masud, E.~S.~Swanson, F.~Akram, and M.~Atif Sultan,
Eur. Phys. J. A \textbf{54}, 127 (2018).

\bibitem{Lu:2016bbk}
Q.~F.~L{\"u}, T.~T.~Pan, Y.~Y.~Wang, E.~Wang, and D.~M.~Li,
Phys. Rev. D \textbf{94}, 074012 (2016).

\bibitem{Geng:2018qrl}
Z.~K.~Geng, T.~Wang, Y.~Jiang, G.~Li, X.~Z.~Tan, and G.~L.~Wang,
Phys. Rev. D \textbf{99}, no.1, 013006 (2019).

\bibitem{Wang:2020zbr}
G.~L.~Wang, T.~F.~Feng, and X.~G.~Wu,
Phys. Rev. D \textbf{101}, no.11, 116011 (2020).

\bibitem{Li:2024dyi}
W.~Li, T.~Wang, T.~F.~Feng, G.~L.~Wang, and C.~H.~Chang,
Phys. Rev. D \textbf{110}, no.7, 076002 (2024).

\end{thebibliography}
\end{document}